\documentclass[aps,nofootinbib,preprintnumbers]{revtex4}
\usepackage{lipsum}
\usepackage{amsfonts}
\usepackage{amsmath}
\usepackage{amssymb}
\usepackage[english]{babel}
\usepackage{graphicx}
\usepackage{epsfig}
\usepackage{bm}
\usepackage{verbatim}
\usepackage[utf8]{inputenc}
\usepackage{multirow}
\usepackage{subfig}
\usepackage{slashed}
\usepackage{xcolor}
\usepackage[colorlinks=true,urlcolor=red,citecolor=red]{hyperref}
\usepackage[font=small]{caption}
\usepackage{float}
\usepackage{blindtext}
\usepackage{bbm}
\usepackage[colorinlistoftodos]{todonotes}
\usepackage[normalem]{ulem}

\newenvironment{Eqnarray}{\arraycolsep 0.14em\begin{eqnarray}}{\end{eqnarray}}
\newcommand{\ba}{\begin{Eqnarray}}
\newcommand{\ea}{\end{Eqnarray}}
\newcommand{\be}{\begin{equation}}
\newcommand{\ee}{\end{equation}}
\newcommand{\bal}{\begin{aligned}}
\newcommand{\eal}{\end{aligned}}
\newcommand{\bea}{\begin{eqnarray}}
\newcommand{\eea}{\end{eqnarray}}
\newcommand{\ben}{\begin{enumerate}}
\newcommand{\een}{\end{enumerate}}
\newcommand{\bit}{\begin{itemize}}
\newcommand{\eit}{\end{itemize}}
\newcommand{\bde}{\begin{widetext}}
\newcommand{\ede}{\end{widetext}}

\def\lsim{\mathrel{\rlap{\lower4pt\hbox{\hskip1pt$\sim$}}
    \raise1pt\hbox{$<$}}}
\def\gsim{\mathrel{\rlap{\lower4pt\hbox{\hskip1pt$\sim$}}
    \raise1pt\hbox{$>$}}}

\def\3211{$\mathrm{SU(3) \otimes SU(2)_L \otimes U(1)_R \otimes U(1)_{B-L}}$ }
\def\321{$\mathrm{SU(3) \otimes SU(2) \otimes U(1)}$ }
\def\422{$\mathrm{SU(4) \otimes SU(2) \otimes SU(2)_R}$ }

\newcommand{\mathsym}[1]{{}}

\definecolor{bostonuniversityred}{rgb}{0.8, 0.0, 0.0}

\input{tcilatex}
\begin{document}

\title{Effective interactions for the SM fermion mass hierarchy and their
possible UV realization.}
\date{\today }
\author{A. E. C\'arcamo Hern\'andez$^{a,b,c}$}
\email{antonio.carcamo@usm.cl}
\author{Diego Restrepo$^{d}$}
\email{restrepo@udea.edu.co}
\author{Ivan Schmidt$^{a,b,c}$\footnote{Iván Schmidt passed away during the revision of this paper. He will be sorely missed.}}
\email{ivan.schmidt@usm.cl}
\author{\'Oscar Zapata$^{d}$}
\email{oalberto.zapata@udea.edu.co}
\affiliation{$^{{a}}$Universidad T\'ecnica Federico Santa Mar\'{\i}a, Casilla 110-V,
Valpara\'{\i}so, Chile\\
$^{{b}}$Centro Cient\'{\i}fico-Tecnol\'ogico de Valpara\'{\i}so, Casilla
110-V, Valpara\'{\i}so, Chile\\
$^{{c}}$Millennium Institute for Subatomic physics at high energy frontier -
SAPHIR, Fernandez Concha 700, Santiago, Chile\\
$^{{d}}$Instituto de F\'{\i}sica, Universidad de Antioquia, Calle 70 No
52-21, Medell\'{\i}n, Colombia }
\date{\today }

\begin{abstract}
We built an extended 2HDM theory with a spontaneously broken $U(1) _{X}$ global symmetry, where the tree level Universal Seesaw Mechanism generates the mass hierarchy of the Standard Model charged fermions and the Zee-Babu mechanism produces tiny active neutrino masses. The third family of SM charged fermions gets tree level masses from Yukawa interactions involving the Higgs doublets $H_1$ (for the top quark) and $H_2$ (for the bottom quark and tau lepton). The model under consideration is consistent with SM fermion masses and mixings, with the muon and electron $g-2$ anomalies and successfully accommodates the constraints arising from charged lepton flavor violation and meson oscillations. The proposed model predicts rates for charged lepton flavor violating decays within the reach of forthcoming experiments.
\end{abstract}
\author{}
\maketitle

\section{Introduction}\label{intro} 
The neutrino masses have been well-established to be exceedingly small \cite{Fukuda:1998mi, Ahmad:2002jz}. A natural explanation for this phenomenon is the tree-level realization of the Weinberg operator~\cite{Weinberg:1979sa} via the type I-III seesaw mechanisms \cite{Minkowski:1977sc, GellMann:1980vs, Yanagida:1979as, Mohapatra:1979ia, Schechter:1980gr, Schechter:1981cv} containing five dimensional operators. 
In order to extend this strategy to all charged matter particles within the Standard Model (SM), the universal seesaw mechanism (USM) \cite{Berezhiani:1983hm, Rajpoot:1987fca, Davidson:1987mh, Davidson:1987mi} has been developed. In the universal seesaw mechanism, heavy charged vector-like fermions, added to the particle spectrum of the SM, mix with the SM charged fermions and yield SM charged fermion masses which will be directly proportional to the product of the two dimensionfull parameters that parametrize their mixings with the SM charged fermions and inversely proportional to the mass scale of the heavy charged vector-like fermions seesaw messengers. This corresponds to a type I seesaw mechanism extended to the charged fermion sector and thus mediated by electrically charged vector-like Dirac fermions. In such universal seesaw realization, the effective Yukawa couplings are directly proportional to a product of two dimensionless couplings, so a small hierarchy in those couplings can yield a quadratically larger hierarchy in the effective couplings. One of the significant advantages of USM is that it predicts the existence of much lighter mediators compared to those involved in the usual seesaws for neutrinos, which in turn can be solved by using either alternative higher-dimensional effective operators or radiative realizations of the Weinberg operator.
Additionally, the quark sector poses further challenges in explaining the observed mixing and mass pattern, necessitating a more elaborate approach that combines higher-dimensional effective operators or incorporates radiative contributions within a extended Higgs sector. 

In this work we invoke to the USM within the framework of a two Higgs
doublet model (2HDM)~\cite{Branco:2011iw}, which can be considered a minimal extension of the SM, with a global horizontal symmetry $U(1)_{X}$ to
generate the masses of the first and second generation of SM
quarks and SM charged leptons. Relying on this symmetry, the third generation charged fermions obtain
their masses thanks to the Yukawa interactions with the Higgs doublets as follows
\begin{align}  \label{eq:tl}
-\mathcal{L}_{d=4}=y_{3}^{\left( u\right) }\overline{q}_{3L}\widetilde{H}%
_{1}\,u_{3R}+y_{3}^{\left( d\right) }\overline{q}_{3L}H_{2}\,d_{3R}+%
\sum_{i=1}^{3}y_{i}^{\left( e\right) }\overline{l}_{iL}H_{2}\,e_{3R}+\text{h.c.}\,,
\end{align}%
where $H_{1}$ and $H_{2}$ are the $SU\left( 2\right) _{L}$ scalar doublets.
Hence that the heaviness of the top
quark with respect to the bottom quark and tau lepton arises from a mild
hierarchy between the vacuum expectation values of $H_{1}$ and $H_{2}$.

In this work we propose a low scale renormalizable model where the set of $d=6$ effective operators responsible for the implementation of the USM that produces the SM charged fermion mass hierarchy are generated at tree level. This is done through
the introduction of several sets of chiral fields transforming under the SM
and $U(1)_{X}$ symmetries. Furthermore, the $d=11$ effective operators that give rise to the tiny active neutrino masses are generated at two-loop level through the Zee-Babu
mechanism. To illustrate the viability of the model we
perform a phenomenological analysis taking into account the current
constraints coming from low energy physics, anomalous magnetic moments of
leptons, charged lepton flavor violation and meson oscillations. The main novelty of our model is thus that it has the implementation of both the USM to generate the SM charged fermion mass hierarchy and the Zee-Babu mechanism for producing the tiny active neutrino masses within the framework of an extended 2HDM theory with moderate particle content and just one extra $U(1)_X$ global symmetry. 

The rest of the paper is organized as follows. In section \ref{sec:model}, we outline the
particle content, the interactions of the model and discuss its implications in SM fermion masses and mixings. The phenomenological consequences of the model concerning $g-2$ muon and electron anomalies as well as charged lepton flavor violation are analyzed in section \ref{gminus2andlfv}. In section \ref{mesons}, we discuss the implications of the model in meson mixings. Finally, we conclude in
section \ref{conclusions}. 

\section{Framework}
\label{sec:model}
To correctly reproduce the mixing and masses pattern in the quark sector~\cite{Xing:2020ijf,Workman:2022ynf}
through USM we further extend the scalar sector with two gauge singlet
fields ($\sigma ,\eta $) and a hyperchargeless weak $SU(2) _{L}$
triplet ($\Delta $), all of them with non zero charges under the $U(1)_{X}$  symmetry. With this setup, one posibility for the the source of the rest of quark
masses are the effective operators of dimension $d=6$
%
\begin{align}
\label{eq:fsl}
  -\mathcal{L}_{d=6}^{\left( q\right) }=&\sum_{n=1}^{2}\left[\gamma _{n}^{\left(
d\right) }\overline{q}_{nL}H_{2}d_{2R} +\gamma_{n}^{\left( u\right)
}\overline{q}_{nL}\widetilde{H}_{2}u_{2R}\right]\frac{\sigma^{\ast }\eta}{\Lambda
^{2}} +\sum_{n=1}^{2}\left[\alpha_{n}^{\left( d\right) }\overline{q}_{nL}\Delta^{\dag }H_{2}d_{1R} + \alpha
_{n}^{\left( u\right) }\overline{q}_{nL}\Delta
\widetilde{H}_{2}u_{1R}\right.  \notag \\
&\left. +\, \alpha_{3n}^{\left( d\right)
}\overline{q}_{nL}\Delta^{\dag} H_{1}d_{3R}+\alpha_{3n}^{\left( u\right)
}\overline{q}_{nL}\Delta^{\dag} \widetilde{H}_{2}u_{3R}\right]\frac{\sigma^{\ast
}}{\Lambda^{2}} + \text{h.c..}
\end{align}
Here $\gamma _{n}^{(d)}, \gamma _{n}^{(u)},\alpha _{n}^{(d)},\alpha
_{n}^{(u)},\alpha _{3n}^{(d)},\alpha _{3n}^{(u)}$ are Yukawa couplings, $%
\Lambda $ signals the energy scale at which new potential fermionic degrees
of freedom enter. 
The terms in the second row are required to generate the mixings with the third family. 
On the other hand, the masses of the electron and muon also arise from the $%
d=6$ operators 
\begin{align}
  \label{eq:fsll}
-\mathcal{L}_{d=6}^{\left( l\right) }=\sum_{i=1}^{3}\gamma _{i}^{\left(
l\right) }\overline{l}_{iL}H_{2}e_{2R}\frac{\sigma ^{\ast }\eta }{\Lambda
^{2}}+\sum_{i=1}^{3}\alpha _{i}^{\left( l\right) }\overline{l}_{iL}\Delta
^{\dag }H_{2}e_{1R}\frac{\sigma ^{\ast }}{\Lambda ^{2}}+\text{h.c.}\,,
\end{align}%
with $\gamma_{i}^{(l)}$ and $\alpha _{i}^{(l)}$ Yukawa couplings.

In our framework, the gauge singlet scalar fields $\sigma$ and $\eta$ acquire vacuum expectation values (VEVs) at an energy scale slightly below the $\Lambda$ scale of the fermionic seesaw mediators. Since these VEVs, $v_\sigma$ and $v_\eta$, closely approach the $\Lambda$ scale, we employ six-dimensional Yukawa operators as opposed to the conventional five-dimensional ones, which are present in the usual type I-III seesaws. This choice aims to provide a more intrinsic rationale for the relatively small masses observed in the first and second generations of SM charged fermions. The hierarchical contrast in VEVs between the triplet ($v_\Delta$) and singlet scalars offers a potential explanation for the observed mass disparity between the first and second generation of SM charged fermions.

At this point there are 10 independent charge equations, leading to 7 free charges. By choosing the set of free $U(1)_X$ charges to be $(\alpha_{H_1},\alpha_{q_3},\alpha_{l},\alpha_{q_{n}},\alpha_{\Delta},\alpha_{\sigma},\alpha_{\eta})$, the complete set of charges allowing all the above $d=4,6$ interaction terms are
\begin{align}
\label{eq:chrgdfer}
  \alpha_{u_1}=&  
  -2 \alpha_{\Delta} +\alpha_{H_1}+\alpha_{q_3}\,,
  & \alpha_{u_2}=& 
  -\alpha_{\Delta} -\alpha_{\eta} +\alpha_{H_1}+\alpha_{q_3}\,,\nonumber\\
   \alpha_{u_3}=& \alpha_{H_1}+\alpha_{q_3}\,,
    &\alpha_{d_1}=&  2 \alpha_{\Delta} -\alpha_{H_1}+2 \alpha_{q_n}-\alpha_{q_3}+2\alpha_{\sigma}\,,\nonumber\\ 
    \alpha_{d_2}=& \alpha_{\Delta} -\alpha_{\eta} -\alpha_{H_1}+2 \alpha_{q_n}-\alpha_{q_3}+2 \alpha_{\sigma}\,,& \alpha_{d_3}=&\alpha_{\Delta} -\alpha_{H_1}+\alpha_{q_n}+\alpha_{\sigma}\,,\nonumber\\
  \alpha_{e_1}=& 2 \alpha_{\Delta}
   -\alpha_{H_1}+\alpha_{l}+\alpha_{q_n}-\alpha_{q_3}+2 \alpha_{\sigma}\,,& \alpha_{e_2}=& \alpha_{\Delta} -\alpha_{\eta} -\alpha_{H_1}+\alpha_{l}+\alpha_{q_n}-\alpha_{q_3}+2 \alpha_{\sigma}\,,&&\nonumber\\
   \alpha_{e_3}=& \alpha_{\Delta} -\alpha_{H_1}+\alpha_{l}+\alpha_{q_n}-\alpha_{q_3}+\alpha_{\sigma}\,, &\alpha_{H_2} =& -\alpha_{\Delta} +\alpha_{H_1}-\alpha_{q_n}+\alpha_{q_3}-\alpha_{\sigma}\,.&&
\end{align}
Here, the three lepton doublets have the same $X$-charge $\alpha_{l}$, and $\alpha_{q_1}=\alpha_{q_2}\equiv \alpha_{q_{n}}$. We have checked with \texttt{Sym2Int}~\cite{Fonseca:2017lem,Fonseca:2019yya} that  the charge assignment only allows the effective operators claimed here.

The neutrino sector deserves a different treatment due to the smallness of their masses. 
On one hand, since the lepton doublets have the same $X$-charge, effective Yukawa operators with right-handed  neutrinos and powers of scalar singlets (or trace of the square of the scalar triplets) are automatically allowed after fixing their common $X$-charge. In absence of effective operators with lepton number violation, only
Dirac neutrinos masses could be realized. 

On the other hand, to get further supression for Majorana neutrino masses one may look for lepton number violation via effective operators of at least dimension $d=7$, such as
\begin{align}
    \mathcal{L}_{\nu}=&\,\sum_{i,j=1}^3\,\alpha_{ij} l_{iL}^a l_{jL}^b H_2^c H_2^d \epsilon_{ab}\epsilon_{cd} \frac{\sigma^{*2}}{\Lambda^2}+\text{h.c.},
\end{align}
which in turn would deliver an additional charge condition,
\begin{align}
\label{eq:sigma}
    \alpha_{\sigma}=\frac{1}{2} \left(-\alpha_{\Delta} +\alpha_{H_1}+\alpha_{l}-\alpha_{q_n}+\alpha_{q_3}\right).
\end{align}
In the present study, we contemplate the interplay of two suppression mechanisms for 
neutrino mass, namely radiative realization of effective operators of high
dimensionality \cite{Bonnet:2012kz,AristizabalSierra:2014wal,Arbelaez:2022ejo}. 
Concretely, we consider the $d=11$
lepton number violating effective operators 
\cite{Babu:2001ex,deGouvea:2007qla,Gargalionis:2020xvt} 
\begin{align}
\label{eq:fsllllll}
-\mathcal{L}_{d=11}^{\left( \nu \right) }=
\frac{1}{\Lambda ^{5}}%
\sum_{i,j,k,s=1}^{3}& \overline{l}_{iL}
l_{jL}^{c} \overline{l}_{kL} l_{sL}^{c} \left[\alpha^{11}_{ijks}\, \overline{%
e_{1R}^{c}}e_{1R}\frac{\Delta^{\ast 2}}{\Lambda ^{2}}
+
\alpha^{22}_{ijks} \overline{e_{2R}^{c}}e_{2R}\frac{%
\eta^{2 }}{\Lambda ^{2}}+\alpha^{33}_{ijks} \overline{e_{3R}^{c}}e_{3R}\frac{%
\sigma ^{2 }}{\Lambda ^{2}}
\right]
+\mathrm{h.c.},
\end{align}%
in such a way that when $\sigma $, $\eta $ and $\Delta$ develop a nonzero VEV neutrino
Majorana masses may be generated at two-loop level through the Zee-Babu
mechanism~\cite{Zee:1985id,Babu:1988ki}. Here $l_{iL}$ and $e_{iR}$ $(i=1,2,3)$ are the SM $SU(2)_L$ leptonic doublets and right handed SM charged leptonic fields, respectively. The charge conditions originated from $d=11$ terms provide 2 additional independent charge conditions, thus reducing the number of free charges to 5: $(\alpha_{\Delta},\alpha_{H_1},\alpha_{q_3},\alpha_{l},\alpha_{q_{12}})$.

\subsection{Ultraviolet realization}
Our aim is to implement the USM to generate the masses of the first and
second generations of SM charged fermions, with the third generation of SM
charged fermions getting their masses via Yukawa interactions with $SU(2)_{L}$ scalar doublets $H_{1}$ (for the top quark)\ and $H_{2}$ (for
the bottom quark and tau lepton). Hence we start from a 2HDM extended by an
extra $U\left( 1\right) _{X}$ symmetry, assumed to be global for simplicity. 
The $U(1)_X$ symmetry is assumed to be softly broken, with the breaking terms in the scalar sector generating a mass for the resulting pseudo-Goldstone bosons.  

The ultraviolet completion of the USM is achieved by introducing the
following set of chiral fermionic fields: 
\begin{equation}
F_{iL},F_{iR},\Psi _{jL},\Psi _{jR},\hspace{1cm}{i=1,2,3,4;\,\,j=1,2.}
\end{equation}%
The fields $F_{i(L,R)}$ are color triplets, with $F_{1,3}$ weak doublets and 
$F_{2,4}$ weak singlets, whereas the fields $\Psi_{j(L,R)}$ are color
singlets, with $\Psi_{1}$ being a weak doublet and $\Psi_{2}$ a weak
singlet. The scalar spectrum is extended by one hyperchargeless $SU(2)_{L}$
scalar triplet $\Delta $, two electrically neutral $\left( \sigma ,\eta
\right) $ and five electrically charged scalars ($\xi,\,\rho,\,\zeta_{1,2,3}$). 
The extra fermion and scalar content with the corresponding electroweak and $U(1)_{X}$ quantum numbers are displayed in Table~\ref{tab:fermoioncharges}.

The electrically neutral gauge singlet scalars and the scalar
triplet together with the charged exotic vector-like fermions are required
for the implementation of the USM that generates the first and second
generation of SM charged fermion masses. The gauge singlet scalar $\sigma $
provides tree level masses to the charged exotic vector like fermions,
whereas the scalar singlet $\eta $ generates mixings between left-handed
charged exotic fermions and the second family of right handed SM fermions.
Moreover, the scalar doublet $H_{2}$, besides generating the
bottom quark and tau lepton masses, also generates mixings between
left-handed charged exotic fermions and the first family of right handed SM
fermions. Furthermore, the mixings between the left-handed SM fermionic
fields and right-handed heavy fermionic seesaw mediators inducing the first
generation of SM charged fermion masses, arise from Yukawa interactions
involving the scalar triplet $\Delta$. 
Besides these scalars, the inclusion of the
electrically charged scalars is necessary for the implementation of the
two-loop level Zee-Babu mechanism that generates the tiny active neutrino
masses.

It is worth stressing that the VEV hierarchy between $v_\Delta$ and  $v_\eta$ and $v_\sigma$ is crucial for explaining the mass hierarchy between the first and second family of SM charged fermions. 
Let us note that, whereas the VEV of the
singlets can be free parameters, $v_\Delta$ should be at most few GeV in order to successfully comply with the constraints arising from the oblique $T$ parameter~\cite{Workman:2022ynf}. Given that our model is non supersymmetric, the aforementioned VEV hierarchy will be unstable under radiative corrections. In order that such VEVs be stable under radiative corrections, we need to impose several sets of Veltmann conditions relating a combination of the quartic scalar couplings that involve a pair of these scalar fields with the remaining ones and the combination of the Yukawa couplings of these scalar fields with the heavy vector-like fermions. Those Veltmann conditions will arise by requiring the cancellation of the quadratically divergent scalar and fermionic contributions, contributions that interfere destructively. Such requirement of tadpole cancellation is an ad hoc condition of viability of our model. In our model we do not have a symmetry responsible for stabilizing the required tadpole cancellation. Introducing such symmetry will require a radical modification of the model structure with all its nice features. This problem can be solved by imbedding our model into a more fundamental setup with additional symmetries protecting the tadpole cancelation. Given that this condition relates the parameters of the fermionic and scalar sectors, two ways of ensuring the stability of the VEVs of the scalar fields of our model under radiative corrections in the whole region of parameter space, will be imbedding our model into a supersymmetric or warped five-dimensional framework. This requires careful studies which are left beyond the scope of the present paper and will be addressed elsewhere.

\begin{table}[t]
\begin{tabular}{|c|c|c|c|c|}
\hline
& $SU(3)_{C}$ & $SU(2)_{L}$ & $U(1)_{Y}$ & $U(1)_{X}$ \\ \hline
$F_{1(L,R)}$ & 3 & 2 & $\frac{1}{6}$ & $\alpha_{F_{1(L,R)}}$ \\ 
$F_{2(L,R)}$ & 3 & 1 & $\frac{2}{3}$ & $\alpha_{F_{2(L,R)}}$ \\ 
$F_{3(L,R)}$ & 3 & 2 & $\frac{1}{6}$ & $\alpha_{F_{3(L,R)}}$ \\ 
$F_{4(L,R)}$ & 3 & 1 & $-\frac{1}{3}$ & $\alpha_{F_{4(L,R)}}$ \\\hline 
  $\Psi _{1(L,R)}$ & 1 & 2 & $-\frac{1}{2}$ & $\alpha_{\Psi _{1(L,R)}}$ \\ 
$\Psi _{2(L,R)}$ & 1 & 1 & -1 & $\alpha_{\Psi _{2(L,R)}}$ \\ \hline
$l$ & 1 & 2 & $-\frac{1}{2}$ & $\alpha_{l}$ \\ 
$e_1$ & 1 & 1 & $-1$ & $\alpha_{e_1}$ \\ 
$e_2$ & 1 & 1 & $-1$ & $\alpha_{e_2}$ \\ 
$e_3$ & 1 & 1 & $-1$ & $\alpha_{e_3}$ \\\hline
\end{tabular}%
\,\,\,\,
\begin{tabular}{|c|c|c|c|c|}
\hline
& $SU(3)_{C}$ & $SU(2)_{L}$ & $U(1)_{Y}$ & $U(1)_{X}$ \\ \hline
$H_{1,2}$ & 1 & 2 & $\frac{1}{2}$ & $\alpha_{H_{1,2}}$ \\ 
$\sigma $ & 1 & 1 & 0 & $\alpha_{\sigma} $ \\ 
$\eta $ & 1 & 1 & 0 & $\alpha_{\eta} $ \\ 
$\Delta $ & 1 & 3 & 0 & $\alpha_{\Delta} $ \\ 
$\xi ^{\pm }$ & 1 & 1 & $\pm 1$ & $\alpha_{\xi} $ \\ 
$\rho ^{\pm \pm }$ & 1 & 1 & $\pm 2$ & $\alpha_{\rho} $ \\ 
$\zeta_{1,2,3}^{\pm \pm }$ & 1 & 1 & $\pm 2$ & $\alpha_{\zeta_{1,2,3}}$ \\\hline
$q_{1,2}$ & 3 & 2 & $\frac{1}{6}$ & $\alpha_{q_{12}}$ \\ 
$q_3$ & 3 & 2 & $\frac{1}{6}$ & $\alpha_{q_3}$ \\ 
$u_1$ & 3 & 1 & $\frac{2}{3}$ & $\alpha_{u_1}$ \\ 
$u_2$ & 3 & 1 & $\frac{2}{3}$ & $\alpha_{u_2}$ \\ 
$u_3$ & 3 & 1 & $\frac{2}{3}$ & $\alpha_{u_3}$ \\
$d_1$ & 3 & 1 & $-\frac{1}{3}$ & $\alpha_{u_1}$ \\ 
$d_2$ & 3 & 1 & $-\frac{1}{3}$ & $\alpha_{u_2}$ \\ 
$d_3$ & 3 & 1 & $-\frac{1}{3}$ & $\alpha_{u_3}$ \\\hline
\end{tabular}%
\caption{Extra fermion and scalar content with the electroweak and $U(1)_{X}$ quantum
numbers. }
\label{tab:fermoioncharges}
\end{table}


By using Eqs.~\eqref{eq:chrgdfer} and~\eqref{eq:sigma}, the ultaviolet completion of the effective model demands that the set of new $U(1)_X$-charges is 
\begin{align}
&\alpha_{\xi} = -2 \alpha_{l},\,\hspace{1cm}\alpha_{\rho}= -4\alpha_{l}, \hspace{1cm}\alpha_{\zeta_1} = -2\alpha_{e_1}=-2 (\alpha_{\Delta} +2 \alpha_{l}),\notag\\
  &  \alpha_{\zeta_2}=-2\alpha_{e_2}=-2 (2 \alpha_{l}-\alpha_{\eta} )-2 (2 \alpha_{l}-\alpha_{\eta} ),\hspace{1cm}\alpha_{\zeta_3}=-2\alpha_{e_3}=-\alpha_{\Delta} + \alpha_{H_1}-3 \alpha_{l}-\alpha_{q_n}+\alpha_{q_3},\,
\end{align}
where the set of free charges is still $(\alpha_{\Delta},\alpha_{H_1},\alpha_{q_3},\alpha_{l},\alpha_{q_{12}})$.  For the extra fermions the $H$-charges are fixed to
\begin{align}
\alpha_{F_{1L}}=&\frac{1}{2} \left(-3 \alpha_{\Delta} +\alpha_{H_1}+\alpha_{l}+\alpha_{q_n}+\alpha_{q_3}\right),\,\,\,\, \alpha_{F_{1R}}=\alpha_{q_n}-\alpha_{\Delta},\,\,\,\, \alpha_{F_{2 L}}=-\alpha_{\Delta} +\alpha_{H_1}+\alpha_{q_3},\,\,\notag\\
\alpha_{F_{2 R}}=&\frac{1}{2} \left(-\alpha_{\Delta} +\alpha_{H_1}-\alpha_{l}+\alpha_{q_n}+\alpha_{q_3}\right), \,\,\,\,
\alpha_{F_{3 L}}=\frac{1}{2} \left(\alpha_{\Delta}+\alpha_{H_1}+\alpha_{l}+\alpha_{q_n}+\alpha_{q_3}\right),\,\,\notag\\
\alpha_{F_{3 R}}=&\alpha_{\Delta} +\alpha_{q_n},\,\,\alpha_{F_{4 L}}=\alpha_{l}+\alpha_{q_n},\,\,\,\,
   \alpha_{F_{4 R}}=\frac{1}{2} \left(\alpha_{\Delta} -\alpha_{H_1}+\alpha_{l}+3 \alpha_{q_n}-\alpha_{q_3}\right),\notag\\
\alpha_{\Psi_{1L}}=&\frac{1}{2} \left(\alpha_{\Delta} +\alpha_{H_1}+3 \alpha_{l}-\alpha_{q_n}+\alpha_{q_3}\right),\,\,\,\,\alpha_{\Psi_{1R}}=\alpha_{\Delta} +\alpha_{l},\,\,\,\, \alpha_{\Psi _{2 L}}=2 \alpha_{l,}\,\, \notag\\
\alpha_{\Psi _{2 R}}=&\frac{1}{2} \left(\alpha_{\Delta} -\alpha_{H_1}+3 \alpha_{l}+\alpha_{q_n}-\alpha_{q_3}\right).
\end{align}
The non-free charges associated with $H_2$ and SM fermions are given by
\begin{align}
&\alpha_{H_2}=-\alpha_l+\alpha_\Delta, \,\,\, \alpha_{u_1}=-2\alpha_\Delta+\alpha_{H_1}+\alpha_{q_3},\,\,\,\alpha_{u_2}=\alpha_l-\alpha_\Delta+2\alpha_{H_1}+2\alpha_{q_3}-\alpha_{q_{12}},\,\,\,\alpha_{u_3}=\alpha_{H_1}+\alpha_{q_3},\nonumber\\
 &   \alpha_{d_1}=2\alpha_{l}-2\alpha_\Delta+\alpha_{H_1}+\alpha_{q_3},\,\,\,\alpha_{d_2}=3\alpha_l-3\alpha_\Delta+2\alpha_{H_1}+2\alpha_{q_3}-\alpha_{q_{12}},\,\,\,\alpha_{d_3}=\alpha_{l}-\alpha_\Delta+\alpha_{q_3},\nonumber\\
  &  \alpha_{e_1}=3\alpha_l-2\alpha_\Delta+\alpha_{H_1}+\alpha_{q_3}-\alpha_{q_{12}},\,\,\,\alpha_{e_2}=4\alpha_l-3\alpha_\Delta+2\alpha_{H_1}+2\alpha_{q_3}-2\alpha_{q_{12}},\,\,\,\alpha_{e_3}=2\alpha_{l}-\alpha_{\Delta}.
\end{align}

With the charge assignment the quark Yukawa interactions are
\begin{align}
-\mathcal{L}_{Y}^{\left( q\right) }& =\sum_{n=1}^{2}x_{n}^{\left( u\right) }%
\overline{q}_{nL}\widetilde{H}_{2}F_{2R}+z_{u}\overline{F}_{2L}\eta
u_{2R}+\sum_{n=1}^{2}x_{n}^{\left( d\right) }\overline{q}%
_{nL}H_{2}F_{4R}+z_{d}\overline{F}_{4L}\eta
d_{2R}+\sum_{n=1}^{2}w_{n}^{\left( d\right) }\overline{q}_{nL}\Delta ^{\dag
}F_{3R}  \notag \\
& +r_{d}\overline{F}_{3L}H_{2}d_{1R}+\sum_{n=1}^{2}w_{n}^{\left( u\right) }%
\overline{q}_{nL}\Delta F_{1R}+r_{u}\overline{F}_{1L}\widetilde{H}%
_{2}u_{1R}+y_{1}^{\left( F\right) }\overline{F}_{1L}\sigma
F_{1R}+y_{2}^{\left( F\right) }\overline{F}_{2L}\sigma F_{2R}  \notag \\
& +y_{3}^{\left( F\right) }\overline{F}_{3L}\sigma F_{3R}+y_{4}^{\left(
F\right) }\overline{F}_{4L}\sigma F_{4R}+k_{u}\overline{F}_{3L}\widetilde{H}%
_{2}u_{3R}+k_{d}\overline{F}_{3L}H_{1}d_{3R}+\mathrm{{h.c.},}
\end{align}
whereas the charged lepton Lagrangian is given by
\begin{align}
-\mathcal{L}_{Y}^{\left( l\right) }& =\sum_{i=1}^{3}x_{i}^{\left( l\right) }%
\overline{l}_{iL}H_{2}\Psi _{2R}+z_{l}\overline{\Psi }_{2L}\eta
e_{2R}+\sum_{i=1}^{3}w_{i}^{\left( l\right) }\overline{l}_{iL}\Delta ^{\dag
}\Psi _{1R}+r_{l}\overline{\Psi }_{1L}H_{2}e_{1R}+\sum_{i=1,2}y_{i}^{\left(
\Psi \right) }\overline{\Psi }_{iL}\sigma \Psi _{iR}+\mathrm{{h.c.}.}
\end{align}
\begin{figure}
\includegraphics[scale=0.3]{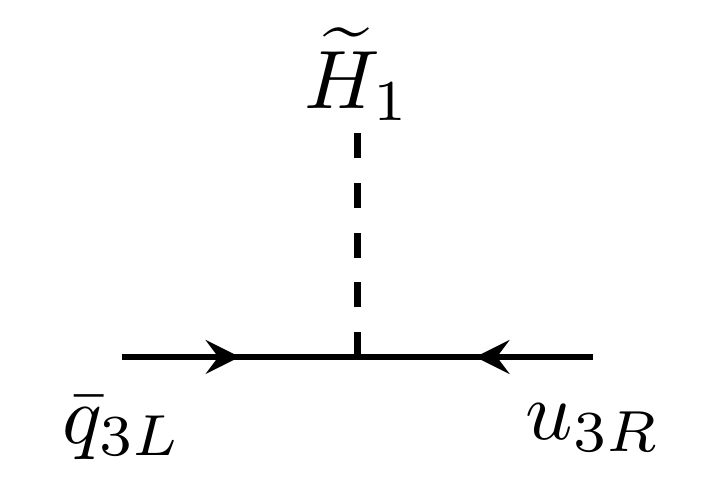} %
\includegraphics[scale=0.3]{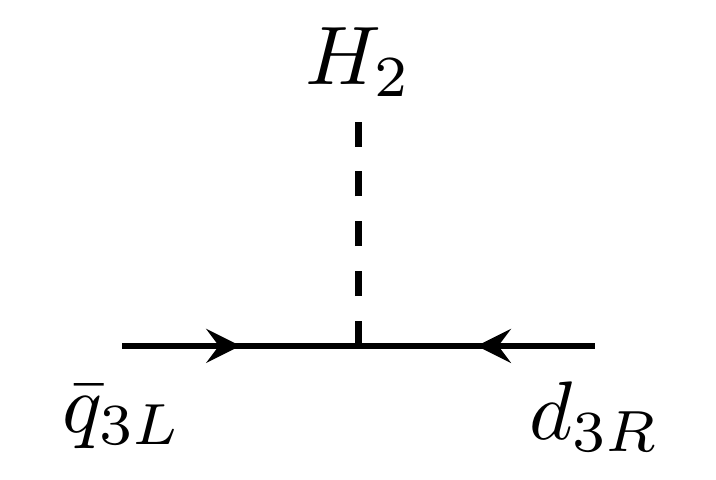}\\
\includegraphics[scale=0.45]{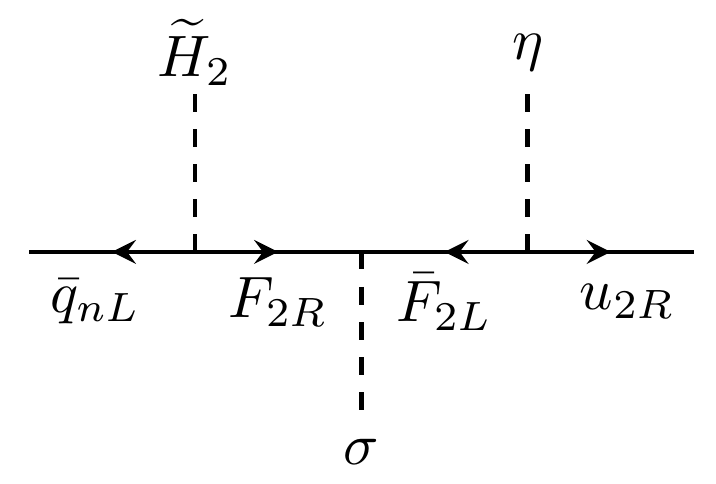}
\includegraphics[scale=0.45]{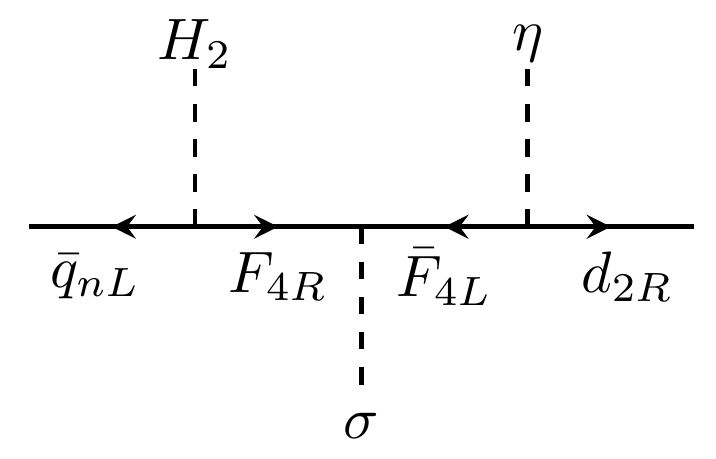}
\includegraphics[scale=0.45]{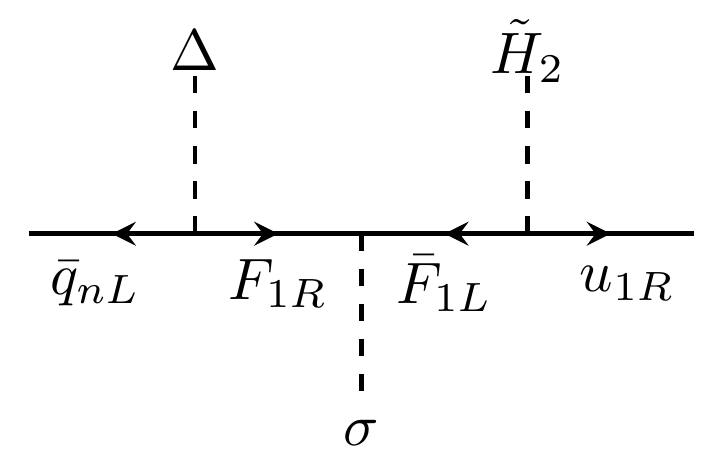}\\
\includegraphics[scale=0.45]{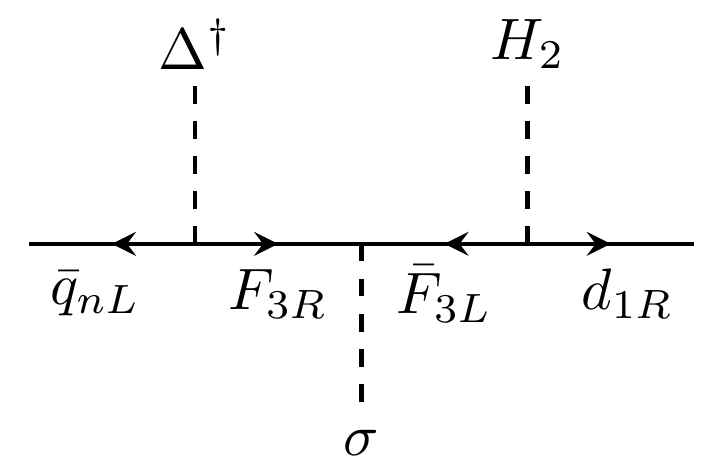}
\includegraphics[scale=0.45]{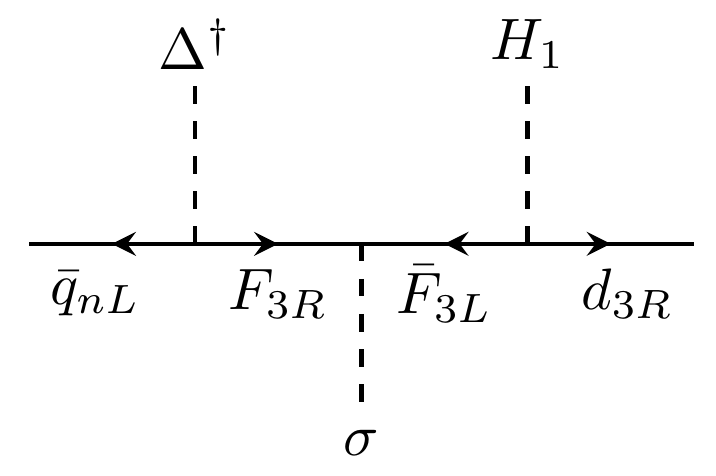} %
\includegraphics[scale=0.45]{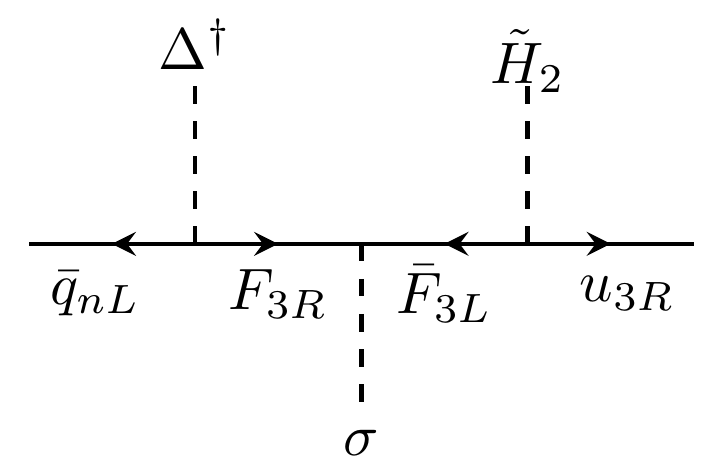}
\caption{Feynman diagrams contributing to the entries of the SM quark mass
matrices. Here, $n=1,2$.}
\label{Diagramsquarks}
\end{figure}
\begin{figure}
\includegraphics[scale=0.3]{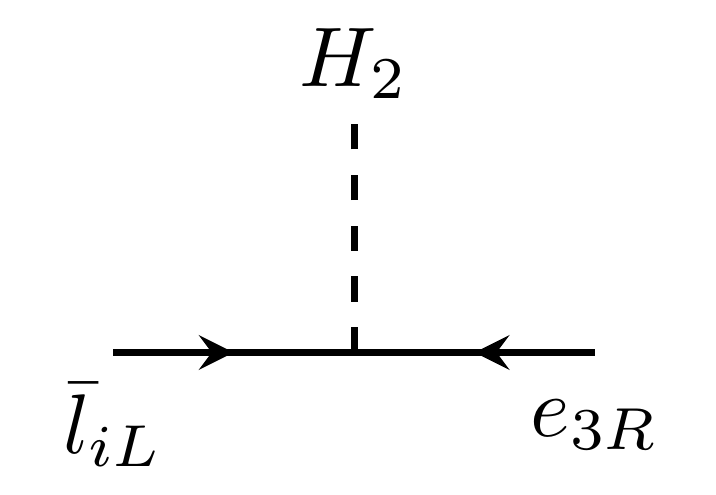}
\includegraphics[scale=0.45]{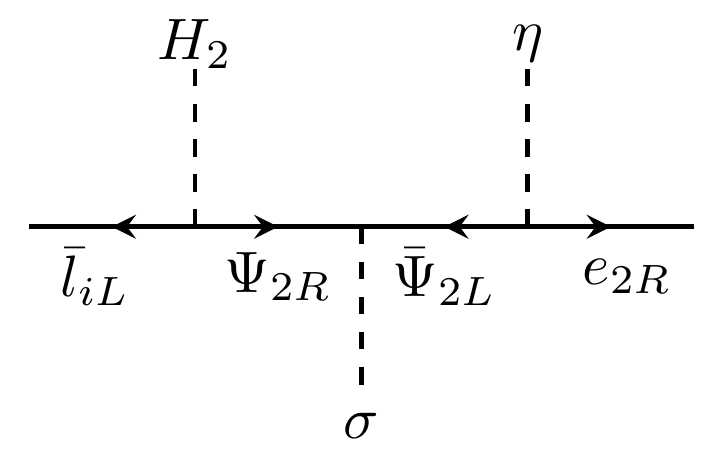}
\includegraphics[scale=0.45]{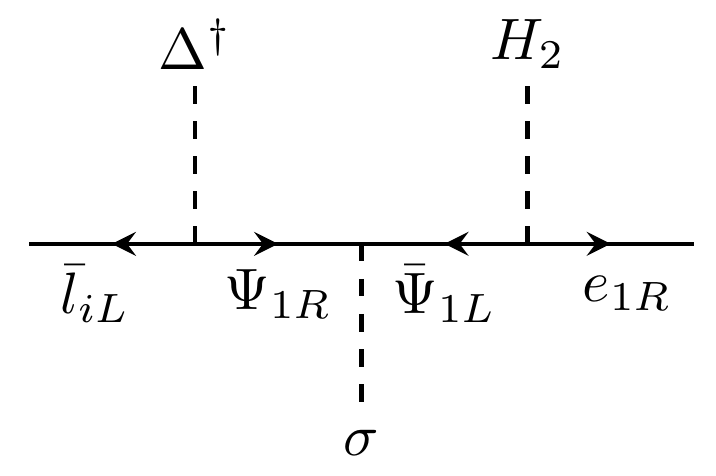}
\caption{Feynman diagrams contributing to the entries of the charged lepton
mass matrix. Here, $i=1,2,3$.}
\label{Diagramschargedleptons}
\end{figure}
%
%
%
%
%
These interactions allow us to construct the seesaw diagrams dispalyed in figures \ref{Diagramsquarks} and \ref{Diagramschargedleptons}, which lead to the following mass matrices for up and down type quarks, 
\begin{eqnarray}
M_{U} &=&\left( 
\begin{array}{cc}
C_{U} & A_{U} \\ 
B_{U} & M_{T}%
\end{array}%
\right) ,\hspace{1cm}\hspace{1cm}A_{U}=\left( 
\begin{array}{ccc}
\frac{w_{1}^{\left( u\right) }v_{\Delta }}{\sqrt{2}} & \frac{x_{1}^{\left(
u\right) }v_{2}}{\sqrt{2}} & \frac{w_{1}^{\left( d\right) }v_{\Delta }}{%
\sqrt{2}} \\ 
\frac{w_{2}^{\left( u\right) }v_{\Delta }}{\sqrt{2}} & \frac{x_{2}^{\left(
u\right) }v_{2}}{\sqrt{2}} & \frac{w_{2}^{\left( d\right) }v_{\Delta }}{%
\sqrt{2}} \\ 
0 & 0 & 0%
\end{array}%
\right) ,\label{MU} \\
M_{D} &=&\left( 
\begin{array}{cc}
C_{D} & A_{D} \\ 
B_{D} & M_{B}%
\end{array}%
\right) ,\hspace{1cm}
A_{D}=\left( 
\begin{array}{ccc}
\frac{w_{1}^{\left( d\right) }v_{\Delta }}{\sqrt{2}} & \frac{x_{1}^{\left(
d\right) }v_{2}}{\sqrt{2}} & \frac{w_{1}^{\left( u\right) }v_{\Delta }}{%
\sqrt{2}} \\ 
\frac{w_{2}^{\left( d\right) }v_{\Delta }}{\sqrt{2}} & \frac{x_{2}^{\left(
d\right) }v_{2}}{\sqrt{2}} & \frac{w_{2}^{\left( u\right) }v_{\Delta }}{%
\sqrt{2}} \\ 
0 & 0 & 0%
\end{array}%
\right) , 
\hspace{1cm}
B_{D} =\left( 
\begin{array}{ccc}
\frac{r_{d}v_{2}}{\sqrt{2}} & 0 & \frac{k_{d}v_{1}}{\sqrt{2}} \\ 
0 & \frac{z_{d}v_{\eta }}{\sqrt{2}} & 0 \\ 
0 & 0 & 0%
\end{array}%
\right),
\label{MD}
\end{eqnarray}%
 and for the charged leptons,
\begin{eqnarray}
M_{l} &=&\left( 
\begin{array}{cc}
C_{E} & A_{E} \\ 
B_{E} & M_{E}%
\end{array}%
\right) ,\hspace{0.5cm}A_{E}=\left( 
\begin{array}{cc}
\frac{w_{1}v_{\Delta }}{\sqrt{2}} & \frac{x_{1}v_{2}}{\sqrt{2}} \\ 
\frac{w_{2}v_{\Delta }}{\sqrt{2}} & \frac{x_{2}v_{2}}{\sqrt{2}} \\ 
\frac{w_{2}v_{\Delta }}{\sqrt{2}} & \frac{x_{3}v_{2}}{\sqrt{2}}%
\end{array}%
\right) , \hspace{0.5cm}
B_{E}=\left( 
\begin{array}{ccc}
\frac{r_{l}v_{2}}{\sqrt{2}} & 0 & 0 \\ 
0 & \frac{z_{l}v_{\eta }}{\sqrt{2}} & 0%
\end{array}%
\right) ,\hspace{0.5cm}
C_{E}=\left( 
\begin{array}{ccc}
0 & 0 & y_{1}^{\left( e\right) }\frac{v_{2}}{\sqrt{2}} \\ 
0 & 0 & y_{2}^{\left( e\right) }\frac{v_{2}}{\sqrt{2}} \\ 
0 & 0 & y_{3}^{\left( e\right) }\frac{v_{2}}{\sqrt{2}}%
\end{array}%
\right).
\end{eqnarray}%
The remaining entries of $M_U, M_D$ and $M_L$ are diagonal matrices with 
\begin{align}
    B_U&={\rm diag} \left(r_uv_2,z_uv_\eta,k_u v_2\right)/\sqrt{2},\hspace{0.5cm} C_U={\rm diag}\left(0,0,1\right)y_3^{(u)}v_1/\sqrt{2},\hspace{0.5cm}  M_T={\rm diag}\left(y_1^{(F)},y_2^{(F)},y_3^{(F)}\right)v_\sigma/\sqrt{2},\nonumber\\
    C_D&={\rm diag}\left(0,0,1\right)y_3^{(d)}v_2/\sqrt{2}, \hspace{0.5cm}  M_B={\rm diag}\left(y_3^{(F)},y_4^{(F)},y_1^{(F)}\right)v_\sigma/\sqrt{2},\hspace{0.5cm} M_E={\rm diag}\left(m_{\Psi_1},m_{\Psi_2}\right).
\end{align}

As follows from the charged fermion Yukawa terms, the heavy vector-like
quarks mix with the SM charged fermions of the first and second generation, thus
triggering a tree-level seesaw mechanism. This in turn generates the first and second generation charged fermion
masses. Futhermore, the masses of the bottom quark and tau lepton are
generated from Yukawa interactions involving a second $SU(2)_{L}$ scalar
doublet which acquires a VEV $v_2$ at the GeV scale, whereas the first scalar
doublet, which gets VEV $v_1$ at the electroweak scale, generates the top quark mass.
Consequently, the SM charged fermion mass matrices are given by: 
\begin{eqnarray}
\widetilde{M}_{U} &=&C_{U}-A_{U}M_{T}^{-1}B_{U}=\left( 
\begin{array}{ccc}
-\frac{r_{u}w_{1}^{\left( u\right) }v_{\Delta }v_{2}}{2m_{F_{1}}} & -\frac{%
z_{u}x_{1}^{\left( u\right) }v_{2}v_{\eta }}{2m_{F_{2}}} & -\frac{%
k_{u}w_{1}^{\left( d\right) }v_{\Delta }v_{2}}{2m_{F_{3}}} \\ 
-\frac{r_{u}w_{2}^{\left( u\right) }v_{\Delta }v_{2}}{2m_{F_{1}}} & -\frac{%
z_{u}x_{2}^{\left( u\right) }v_{2}v_{\eta }}{2m_{F_{2}}} & -\frac{%
k_{u}w_{2}^{\left( d\right) }v_{\Delta }v_{2}}{2m_{F_{3}}} \\ 
0 & 0 & \frac{y_{3}^{\left( u\right) }v_{1}}{\sqrt{2}}%
\end{array}%
\right) =\left( 
\begin{array}{ccc}
C_{u}^{\left( 1\right) } & C_{u}^{\left( 2\right) } & C_{u}^{\left( 3\right)
} \\ 
C_{u}^{\left( 4\right) } & C_{u}^{\left( 5\right) } & C_{u}^{\left( 6\right)
} \\ 
0 & 0 & C_{u}^{\left( 7\right) }%
\end{array}%
\right) ,  \notag \\
\widetilde{M}_{D} &=&C_{D}-A_{D}M_{B}^{-1}B_{D}=\left( 
\begin{array}{ccc}
-\frac{r_{d}w_{1}^{\left( d\right) }v_{\Delta }v_{2}}{2m_{F_{3}}} & -\frac{%
z_{d}x_{1}^{\left( d\right) }v_{2}v_{\eta }}{2m_{F_{4}}} & -\frac{%
k_{d}w_{1}^{\left( d\right) }v_{\Delta }v_{1}}{2m_{F_{3}}} \\ 
-\frac{r_{d}w_{2}^{\left( d\right) }v_{\Delta }v_{2}}{2m_{F_{3}}} & -\frac{%
z_{d}x_{2}^{\left( d\right) }v_{2}v_{\eta }}{2m_{F_{4}}} & -\frac{%
k_{d}w_{2}^{\left( d\right) }v_{\Delta }v_{1}}{2m_{F_{3}}} \\ 
0 & 0 & \frac{y_{3}^{\left( d\right) }v_{2}}{\sqrt{2}}%
\end{array}%
\right) =\left( 
\begin{array}{ccc}
C_{d}^{\left( 1\right) } & C_{d}^{\left( 2\right) } & C_{d}^{\left( 3\right)
} \\ 
C_{d}^{\left( 4\right) } & C_{d}^{\left( 5\right) } & C_{d}^{\left( 6\right)
} \\ 
0 & 0 & C_{d}^{\left( 7\right) }%
\end{array}%
\right) ,  \notag \\
\widetilde{M}_{l} &=&C_{E}-A_{E}M_{E}^{-1}B_{E}=\left( 
\begin{array}{ccc}
-\frac{r_{l}w_{1}^{\left( l\right) }v_{\Delta }v_{2}}{2m_{\Psi _{1}}} & -%
\frac{z_{l}x_{1}^{\left( l\right) }v_{2}v_{\eta }}{2m_{\Psi _{2}}} & \frac{%
y_{1}^{\left( e\right) }v_{2}}{\sqrt{2}} \\ 
-\frac{r_{l}w_{2}^{\left( l\right) }v_{\Delta }v_{2}}{2m_{\Psi _{1}}} & -%
\frac{z_{l}x_{2}^{\left( l\right) }v_{2}v_{\eta }}{2m_{\Psi _{2}}} & \frac{%
y_{2}^{\left( e\right) }v_{2}}{\sqrt{2}} \\ 
-\frac{r_{l}w_{3}^{\left( l\right) }v_{\Delta }v_{2}}{2m_{\Psi _{1}}} & -%
\frac{z_{l}x_{3}^{\left( l\right) }v_{2}v_{\eta }}{2m_{\Psi _{2}}} & \frac{%
y_{3}^{\left( e\right) }v_{2}}{\sqrt{2}}%
\end{array}%
\right) .  \label{Mflow}
\end{eqnarray}
On the other hand, despite there is a gap of about one order of magnitude between the masses of the top quark and seesaw messengers, the charged vector-like fermions can be integrated out since they trigger an universal seesaw mechanism that yields the masses of the first and second families of the SM charged fermions, values that differ by several orders of magnitude with the seesaw messenger masses. It is worth mentioning that the masses of the third generation of SM charged fermions are generated from Yukawa interactions with the $SU(2)$ scalar doublets $H_1$ (for the top quark) and $H_2$ (for the bottom quark and tau lepton).

The experimental values for the quark masses ~\cite{Xing:2020ijf} and the
CKM parameters ~\cite{Workman:2022ynf} \footnote{%
These are given by 
\begin{align*}
& m_{u}^{\text{exp}}(M_{Z})=1.24\pm 0.22\,\text{MeV}\;,m_{c}^{\text{exp}%
}(M_{Z})=0.626\pm 0.020\,\text{GeV}\;,m_{t}^{\text{exp}}(M_{Z})=172.9\pm
0.04\,\text{GeV}\;, \\
& m_{d}^{\text{exp}}(M_{Z})=2.69\pm 0.19\,\text{MeV}\;,m_{s}^{\text{exp}%
}(M_{Z})=53.5\pm 4.6\,\text{MeV}\;,m_{b}^{\text{exp}}(M_{Z})=2.86\pm 0.03\,%
\text{GeV}\;; \\
& |\mathbf{V}_{12}^{\text{exp}}|=0.22452\pm 0.00044\;,|\mathbf{V}_{23}^{%
\text{exp}}|=0.04214\pm 0.00076\;,|\mathbf{V}_{13}^{\text{exp}}|=0.00365\pm
0.00012\;, \\
& J_{q}^{\text{exp}}=(3.18\pm 0.15)\times 10^{-5}\;.
\end{align*}%
} can be successfully accommodated for the following benchmark point: 
\begin{eqnarray}
x_{1}^{\left( u\right) } &\simeq &-0.621,\hspace{1cm}x_{2}^{\left(
u\right) }\simeq -0.137,\hspace{1cm}w_{1}^{\left( u\right) }\simeq
0.180-0.740i,\hspace{1cm}w_{2}^{\left( u\right) }\simeq 2.153 + 0.964i, 
\notag \\
k_{u}=r_{u}=z_{u} &=&1.000,\hspace{1cm}y_{3}^{\left( u\right) }\simeq 1.004,\hspace{1cm}v_{2}=6
\mbox{GeV},\hspace{1cm}v_{\Delta }=1\mbox{GeV},\hspace{1cm}v_{\eta }=1%
\mbox{TeV},  \notag \\
x_{1}^{\left( d\right) } &\simeq &0.083+0.056i,\hspace{1cm}x_{2}^{\left( d\right)
}\simeq 0.041,\hspace{1cm}w_{1}^{\left( d\right) }\simeq -2.855,%
\hspace{1cm}w_{2}^{\left( d\right) }\simeq -0.433+0.167i,  \notag \\
k_{d} &\simeq &0.763,\hspace{1cm}r_{d}\simeq 3.436,\hspace{1cm}%
z_{d}\simeq 0.742,\hspace{1cm}y_{3}^{\left( d\right) }\simeq 0.706,\hspace{1cm%
}v_{1}=246\mbox{GeV},  \notag \\
m_{F_{1}} &\simeq &5.6\mbox{TeV},\hspace{1cm}m_{F_{2}}\simeq 2.8\mbox{TeV},%
\hspace{1cm}m_{F_{3}}\simeq 2.3\mbox{TeV},\hspace{1cm}m_{F_{4}}\simeq 4.1%
\mbox{TeV}, 
\label{bpf}
\end{eqnarray}
By performing a numerical diagonalization of the full up and down type quark mass matrices given in Eqs. (\ref{MU}) and (\ref{MD}), we find the absolute values of the entries of the full $6\times 6$ CKM quark mixing matrix $V_{CKM}=\left(V_L^{(U)}\right)^{\dagger}V_L^{(D)}$ are:
\begin{eqnarray}
\bigl|V_{CKM}\bigr|=\left(
\begin{array}{cccccc}
 9.7452\times 10^{-1} & 2.2428\times 10^{-1} & 3.6619\times 10^{-3} & 3.7378\times 10^{-6} & 2.34\times 10^{-5} & 3.5355\times 10^{-6} \\
 2.2412\times 10^{-1} & 9.7366\times 10^{-1} & 4.1978\times 10^{-2} & 9.1215\times 10^{-4} & 9.1527\times 10^{-4} & 9.1498\times 10^{-4} \\
 9.3526\times 10^{-3} & 4.1087\times 10^{-2} & 9.9911\times 10^{-1} & 6.7015\times 10^{-5} & 1.8023\times 10^{-8} & 1.2596\times 10^{-4} \\
 2.0499\times 10^{-4} & 8.963\times 10^{-4} & 8.7684\times 10^{-5} & 8.3962\times 10^{-7} & 2.1253\times 10^{-6} & 1. \\
 1.8554\times 10^{-4} & 8.9575\times 10^{-4} & 3.8414\times 10^{-5} & 8.8554\times 10^{-7} & 1. & 1.2974\times 10^{-6} \\
 2.0394\times 10^{-4} & 8.9111\times 10^{-4} & 2.9054\times 10^{-5} & 1. & 5.9522\times 10^{-8} & 3.4306\times 10^{-9} \\
\end{array}
\right),
\label{VCKM}
\end{eqnarray}
from which it follows that the $3\times 3$ CKM quark mixing matrix $K_{CKM}$ is non unitary thanks to the mixings of the SM quark fields with the heavy non SM vector-like quarks. On the other hand, the full $6\times 6$ CKM quark mixing matrix $V_{CKM}=\left(V_L^{(U)}\right)^{\dagger}V_L^{(D)}$ is unitary, which implies that: 
\begin{equation}
\sum_{k=1}^{6}\left(V_{CKM}\right)_{ik}\left(\left(V_{CKM}\right)^{\dagger}\right)_{kj}=\delta_{ij},\hspace{1.5cm}\sum_{k=1}^{6}\left(V_{CKM}\right)_{ki}\left(\left(V_{CKM}\right)^{\dagger}\right)_{jk}=\delta_{ij}.
\label{UnitarityCKM}
\end{equation}
Then, as a consequence of the unitarity of the $6\times 6$ CKM quark mixing matrix $V_{CKM}$, its rows and columns are orthonormal, and one has, for instance: 
\begin{equation}
V_{ud}V_{ub}^{*}+V_{cd}V_{cb}^{*}+V_{td}V_{tb}^{*}+\sum_{i=1}^{3}V_{T_id}V_{T_ib}^{*}=0,
\label{orthogonalitycondition}
\end{equation}
where $V=V_{CKM}$ and $\bigl|\sum_{i=1}^{3}V_{T_id}V_{T_ib}^{*}\bigr|\sim\mathcal{O}\left(10^{-8}\right)$, as follows from Eq. (\ref{VCKM}). This shows that the unitarity triangle with sides $\bigl|V_{ud}V_{ub}^{*}\bigr|\sim\mathcal{O}\left(10^{-3}\right)$, $\bigl|V_{cd}V_{cb}^{*}\bigr|\sim\mathcal{O}\left(10^{-2}\right)$ and $\bigl|V_{td}V_{tb}^{*}\bigr|\sim\mathcal{O}\left(10^{-2}\right)$ is closed with high accuracy, since the corresponding error is of the order of 
\begin{equation}
\biggl|\frac{\sum_{i=1}^{3}V_{T_id}V_{T_ib}^{*}}{V_{ud}V_{ub}^{*}}\biggr|\sim\mathcal{O}(10^{-5}).    
\end{equation}
Moreover, the non unitary $3\times 3$ CKM quark mixing matrix $K_{CKM}$ corresponding to the upper left $3\times 3$ block of $V_{CKM}$ can be written as follows:
\begin{equation}
K_{CKM}=\left(1_{3\times 3}-\epsilon\right)U_{CKM},    
\end{equation}
where $U_{CKM}$ is a unitary matrix and $\epsilon$ is an Hermitian matrix which parametrizes the departure of $K_{CKM}$ from unitarity. The entries of the $\epsilon$ matrix satisfy:
\begin{equation}
\bigl|\epsilon\bigr|=\left(
\begin{array}{ccc}
 2.8703\times 10^{-10} & 1.3341\times 10^{-8} & 3.4768\times 10^{-10} \\
 1.3341\times 10^{-8} & 1.2535\times 10^{-6} & 8.8181\times 10^{-8} \\
 3.4768\times 10^{-10} & 8.8181\times 10^{-8} & 1.0179\times 10^{-8} \\
\end{array}
\right).
\label{ep}
\end{equation}
Then, as indicated by Eq. (\ref{ep}), the departure from unitarity of the $3\times 3$ CKM quark mixing matrix $K_{CKM}$ is of the order of $10^{-6}$, which is below the corresponding experimental constraints \cite{Belfatto:2019swo,ParticleDataGroup:2024cfk}. For instance, as shown in Eq. (\ref{VCKM}) our model successfully complies with the experimental upper bound on the CKM deviation from the unitarity \cite{Belfatto:2019swo,ParticleDataGroup:2024cfk}:
\begin{equation}
\Delta_{\func{CKM}} = \sqrt{1-V_{ud}^{2}-V_{us}^{2}-V_{ub}^{2}} < 0.04\pm 0.01.
\label{eqn:CKM_non_unitarity}
\end{equation}
Such unitarity deviations give rise to flavor changing neutral processes, such as for instance rare top quark decays $t\rightarrow Z\tilde{u}_n$ and $t\rightarrow h\tilde{u}_n$ ($n=1,2$), where $\tilde{u}_1$ and $\tilde{u}_2$ correspond to up and charm quarks, respectively. The branching ratios for the flavor violating top quark decays $t\rightarrow Z\tilde{u}_n$ and $t\rightarrow h\tilde{u}_n$ ($n=1,2$) are respectively given by \cite{Alves:2023ufm,Branco:2021vhs}:
\begin{equation}
Br\left(t\rightarrow Z\tilde{u}_n\right)=\frac{1-3r_Z^4+2r_Z^6}{1-3r_W^4+2r_W^6}\frac{\bigl|\Xi_{n3}^{(u)}\bigr|^2}{2\bigl|V_{tb}\bigr|^2},\hspace{1cm}Br\left(t\rightarrow h\tilde{u}_n\right)=\frac{\left(1-r_h\right)^2}{1-3r_W^4+2r_W^6}\frac{\cos^2\alpha\bigl|\Xi_{n3}^{(u)}\bigr|^2}{2\bigl|V_{tb}\bigr|^2},   
\end{equation}
where $\alpha$ is a mixing angle in the low energy CP even scalar sector and the dimensionless quantities $r_W$, $r_W$, $r_t$ and $\Xi_{n3}^{(u)}$ are defined as:
\begin{equation}
r_W=\frac{m_W}{m_t},\hspace{1cm}r_Z=\frac{m_Z}{m_t},\hspace{1cm}r_h=\frac{m_h}{m_t},\hspace{1cm}\Xi_{n3}^{(u)}= -\sum_{k=6}^{4}\left(V_L^{(U)}\right)_{nk}\left(\left(V_L^{(U)}\right)^{\dagger}\right)_{k3},\hspace{1cm}n=1,2
\end{equation}
Consequently, for the benchmark point described above, we find $Br\left(t\rightarrow Zc\right)\sim\mathcal{O}\left(10^{-16}\right)$, $Br\left(t\rightarrow Zu\right)\sim \mathcal{O}\left(10^{-16}\right)$, $Br\left(t\rightarrow hc\right)\sim\mathcal{O}\left(10^{-17}\right)$, $Br\left(t\rightarrow hu\right)\sim \mathcal{O}\left(10^{-15}\right)$, which is well below the current experimental limits $Br\left(t\rightarrow Zu\right)<6.2\times 10^{-5}$, $Br\left(t\rightarrow Zc\right)<1.3\times 10^{-4}$ \cite{ATLAS:2023qzr}, $Br\left(t\rightarrow hu\right)<2.6\times 10^{-4}$ and $Br\left(t\rightarrow hc\right)<3.4\times 10^{-4}$~\cite{ATLAS:2024mih}.

\begin{figure}
\includegraphics[scale=0.6]{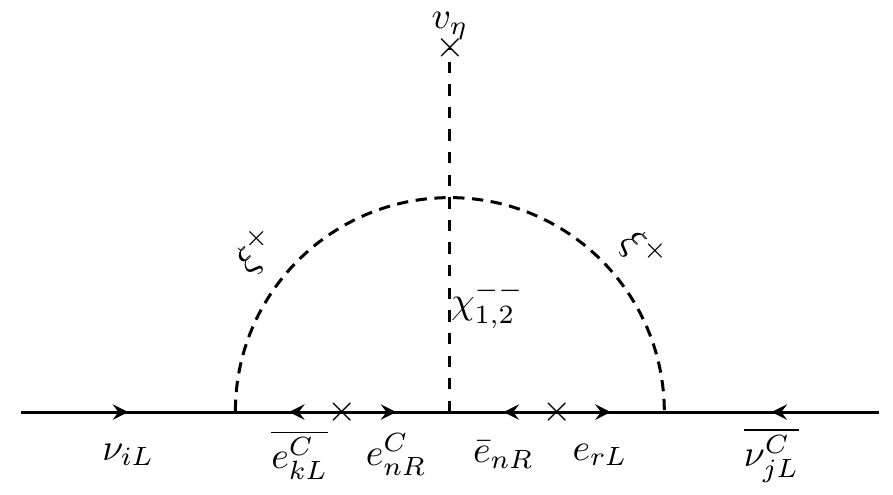}
\caption{Feynman diagrams contributing to the entries of the neutrino mass
matrix. Here, $i,j,k,r=1,2,3$ and $n=1,3$.}
\label{DiagramNeutrinos}
\end{figure}

On the other hand, tiny neutrino masses arise from the interplay of the Yukawa interactions 
\begin{equation}
-\mathcal{L}_{Y}^{\left( \nu \right) }=\sum_{i,j=1}^{3}\kappa
_{ij}\overline{l_{iL}^{c}}l_{jL}\xi ^{+}+\sum_{i=1}^{3}\gamma _{i}\overline{e_{iR}^{c}}e_{iR}\zeta_i^{++}+\mathrm{{h.c.}\,,}
\end{equation}
and the interactions from the scalar potential 
\begin{equation}
\mathcal{V}\supset 
\left[\lambda _{7}\zeta_1^{--}\Delta^{\dagger 2}+\lambda _{8}\zeta_2^{--}\eta^{2}+\lambda _{9}\zeta_3^{--}\sigma^{2}+\mu_{\xi\rho}\xi^{-}\xi^{-}\right]\rho^{++}+\text{h.c.}\,.
\end{equation}
The resulting mass matrix at two loop level for the light active neutrinos (see Fig. \ref{DiagramNeutrinos})
takes the form 
\begin{equation}
M_{\nu }=\sum_{r=1}^{4}\sum_{k=1}^{4}\frac{\mu _{\xi \rho }\left( R_{CC}\right) _{4r}}{48\pi ^{2}m_{k}^{2}}\kappa 
\widetilde{M}_{l}G_{k}^{\dagger }\widetilde{M}_{l}^{T}\kappa ^{T}J\left( 
\frac{m_{\chi _{k}^{--}}^{2}}{m_{\xi ^{+}}^{2}}\right) ,\hspace{1cm}%
m_{k}=\max \left( m_{\chi _{k}^{--}},m_{\xi ^{+}}\right) ,
\end{equation}%
where 
\begin{eqnarray}
J\left( \varkappa \right)  &=&\left\{ 
\begin{array}{l}
1+\frac{3}{\pi ^{2}}\left( \ln ^{2}\varkappa -1\right),\,\, \text{for}\hspace{%
0.2cm}\varkappa >>1 \\ 
\\ 
1,\,\,\text{for}\hspace{0.2cm}\varkappa \rightarrow 0.%
\end{array}%
\right. ,\hspace{1cm}\left( 
\begin{array}{c}
\chi _{1}^{\pm \pm } \\ 
\chi _{2}^{\pm \pm } \\ 
\chi _{3}^{\pm \pm }\\
\chi _{4}^{\pm \pm }
\end{array}%
\right) =R_{CC}^{T}\left( 
\begin{array}{c}
\zeta _{1}^{\pm \pm } \\ 
\zeta _{2}^{\pm \pm } \\
\zeta _{3}^{\pm \pm } \\
\rho ^{\pm \pm }%
\end{array}%
\right), 
\end{eqnarray}
and $G_k={\rm diag}\left(\gamma_1(R_{CC})_{1k}, \gamma_2(R_{CC})_{2k}, \gamma_3(R_{CC})_{3k}\right)$, $k=1,2,3,4$.
Let us note that the scalar interactions impose three additional charge conditions, which determine $\zeta_1,\zeta_2,\xi$ , while maintaining the same set of free charges. For illustration purposes we provide the set of non-free charges for the specific choice of free charges $(\alpha_{\Delta},\alpha_{H_1},\alpha_{q_3},\alpha_{l},\alpha_{q_{12}})=(1,0,1/3,1/3,1/6)$:
\begin{align}\nonumber
&    \alpha_{u_1}=-\frac{5}{3},\,\,\alpha_{u_2}=-\frac{1}{6},\,\,\alpha_{u_3}=\frac{1}{3},\,\,\alpha_{d_1}=-1,\,\,\alpha_{d_2}=-\frac{3}{2},\,\,\alpha_{d_3}=-\frac{1}{3},\,\,\alpha_{e_1=}-\frac{5}{6},\,\,\alpha_{e_2}=-\frac{4}{3},\,\,\alpha_{e_3}=-\frac{1}{3},\,\,\alpha_{H_2}=\frac{2}{3},\,\,\\\nonumber
&\alpha_{\sigma}=-\frac{3}{2},\,\,\alpha_{\eta}=-\frac{1}{2},\,\,\alpha_{\xi}=-\frac{2}{3},\,\,\alpha_{\zeta_1}=\frac{5}{3},\,\,\alpha_{\zeta_2}=\frac{2}{3},\,\,\alpha_{F_{1L}}=-\frac{7}{3},\,\,\alpha_{F_{1R}}=-\frac{5}{6},\,\,\alpha_{F_{2L}}=-\frac{2}{3},\,\,\alpha_{F_{2R}}=\frac{5}{6},\,\,\alpha_{F_{3L}}=-\frac{1}{3},\\
&\alpha_{F_{3R}}=\frac{7}{6},\,\,\alpha_{F_{4L}}=-2,\,\,\alpha_{F_{4R}}=-\frac{1}{2},\,\,\alpha_{\Psi_{1L}}=-\frac{1}{6},\,\,\alpha_{\Psi_{1R}}=\frac{4}{3},\,\,\alpha_{\Psi_{2L}}=-\frac{11}{6},\,\,\alpha_{\Psi_{2R}}=-\frac{1}{3},\,\,\alpha_{\rho}=-\frac{4}{3}\,.
\end{align}

We have conducted a comprehensive numerical analysis to validate the agreement between the measured values of Standard Model charged fermion masses, neutrino mass squared splittings, quark and leptonic mixing parameters, and the CKM and Dirac leptonic CP phases. Our results indicate a successful reproduction of these observables within the $3\sigma$ level within well-defined regions of the parameter space. This accomplishment is achieved while accommodating electrically charged and doubly charged scalar masses at the scale of $1-10$ TeV. 
It is important to note that the Zee-Babu model encounters significant challenges when confronted with neutrino oscillation data~\cite{Herrero-Garcia:2014hfa, Irie:2021obo}.


Finally to close this section we provide a brief discussion of collider signatures of the charged vector like fermions of our proposed model. Given that we are considering the spontaneous breaking of the $U(1)_X$ global symmetry takes place at an scale of few TeV, these exotic fermions obtain TeV scale masses after $U(1)_X$ symmetry breaking. Thus, the charged vector like fermions of our model can be accessible at collider experiments. The charged vector like quarks can be pair produced by Drell-Yan and gluon fusion processes mediated by gauge bosons and gluons, respectively. On the other hand, the charged vector like leptons can be pair produced by Drell Yan mechanism. These charged vector like fermions can decay into a SM charged fermions and one of the scalar fields of the model. Thus, the observation of an excess of events with respect to the SM background in the dijet and opossite sign dileptons final states at the LHC, can be a signal of support of this model.

\section{Charged leptons phenomenology}
\label{gminus2andlfv} 
In this section we will analyze the implications of our model in charged
lepton flavor violation as well as in the muon and electron anomalous
magnetic moments. 

Despite considerable experimental endeavors aimed at detecting signals of lepton rare decays,  evidence of such processes remains elusive \cite{Lindner:2016bgg, Calibbi:2017uvl}. It is worth noting that these experimental pursuits have not only achieved great sensitivity~\cite{Adam:2013mnn,Bellgardt:1987du} but are also on the brink of achieving substantial improvements in the near future, in some cases even by several orders of magnitude~\cite{Baldini:2013ke,Blondel:2013ia}. In contrast, the most recent experimental result for $(g-2)_\mu$~\cite{Muong-2:2023cdq} reported by the Muon $g-2$ Collaboration improves its previous result
by more than a factor of two \cite{Muong-2:2021ojo}, potentially exhibiting deviations from the prediction of the Standard Model~\cite{Aoyama:2020ynm}.

In the present model, the branching ratio for the $l_{i}\rightarrow l_{j}\gamma $
decay receives one loop level contributions from electrically charged and
doubly charged scalars, taking the form \cite{Lindner:2016bgg} 
\begin{equation}
Br\left( l_{i}\rightarrow l_{j}\gamma \right) =\frac{\alpha _{\text{em}}}{%
48\pi G_{F}^{2}}\left( \sum_{s=1}^{3}\left\vert \frac{\left( \left(
R_{C}\right) _{3s}\right) ^{2}\left( \kappa ^{\dagger }\kappa \right) _{ji}}{%
m_{\varphi _{k}^{\pm }}^{2}}\right\vert ^{2}+16\sum_{k=1}^{3}\left\vert 
\frac{\left( G_{k}^{\dagger }G_{k}\right) _{ji}}{m_{\chi _{k}^{\pm \pm }}^{2}%
}\right\vert ^{2}\right) .
\end{equation}%
In order to simplify our analysis, we consider a simplified benchmark
scenario corresponding to the alignment limit, where the $H_{2R}^{0}$, $\eta
_{R}$, $\Delta _{R}^{0}$ and $\sigma _{R}$ do not mix with the neutral CP
even part of $H_{1}$, i.e., $H_{1R}^{0}$. Furthermore, we assume that $%
\sigma $ does not mix with the remaining scalar fields. In that scenario $%
H_{1R}^{0}$ is identified with the $126$ GeV SM like Higgs boson. Moreover, the heavy neutral CP even $H_{2R}^{0}$, $\eta _{R}$, $\Delta
_{R}^{0}$, neutral CP odd $H_{2I}^{0}$, $\eta _{I}$, $\Delta _{I}^{0}$,
electrically charged scalars $H_{2}^{\pm }$, $\xi ^{\pm }$\ and $\Delta
_{R}^{\pm }$ and doubly charged scalars (in the interaction basis) relevant
for the $g-2$ anomalies and charged lepton flavor violating processes are
related with the corresponding scalars in the mass basis by the following
relations: 
\begin{equation}
\left( 
\begin{array}{c}
H_{2R}^{0} \\ 
\eta _{R} \\ 
\Delta _{R}^{0}%
\end{array}%
\right) =R_{H}\left( 
\begin{array}{c}
S_{1} \\ 
S_{2} \\ 
S_{3}%
\end{array}%
\right) ,\hspace{0.7cm}\left( 
\begin{array}{c}
H_{2I}^{0} \\ 
\eta _{I} \\ 
\Delta _{I}^{0}%
\end{array}%
\right) =R_{A}\left( 
\begin{array}{c}
A_{1} \\ 
A_{2} \\ 
A_{3}%
\end{array}%
\right) ,\hspace{0.7cm}\left( 
\begin{array}{c}
H_{2}^{\pm } \\ 
\xi ^{\pm } \\ 
\Delta _{R}^{\pm }%
\end{array}%
\right) =R_{C}\left( 
\begin{array}{c}
\varphi _{1}^{\pm } \\ 
\varphi _{2}^{\pm } \\ 
\varphi _{3}^{\pm }%
\end{array}%
\right) ,\hspace{0.7cm}\left( 
\begin{array}{c}
\zeta _{1}^{\pm \pm } \\ 
\zeta _{2}^{\pm \pm } \\ 
\rho ^{\pm \pm }%
\end{array}%
\right) =R_{CC}\left( 
\begin{array}{c}
\chi _{1}^{\pm \pm } \\ 
\chi _{2}^{\pm \pm } \\ 
\chi _{3}^{\pm \pm }%
\end{array}%
\right) 
\end{equation}%
where $R_{H}$, $R_{A}$, $R_{C}$ and $R_{CC}$ are real orthogonal $3\times 3$
rotation matrices, respectively. 
It is worth noticing that the spontaneous breaking of the global symmetry leads to the emergence of a Goldstone boson, which corresponds to the imaginary part of $\sigma$. As a singlet, this Goldstone boson has minimal impact on the phenomenology of the model. Additionally, if the global symmetry is softly broken, $\sigma_I$ will acquire mass from the soft-breaking mass terms in the scalar potential, once again having little effect on the phenomenology. 

As consequence of the softy broken global $U(1)_X$ symmetry, domain walls (DW) will be formed~\cite{Vilenkin:1981zs,Kibble:1982dd,Vilenkin:1984ib}. After the formation process, they may potentially dominate the universe's energy density even after the light nuclei form, thus spoiling successful predictions of BBN. To avoid this, DW must become, for instance, unstable, assuming that the DW formation process takes place after inflation (see e.g. Ref.~\cite{Lazarides:2018aev} for alternative strategies to prevent the DW formation). 
These soft-breaking scalar mass terms can give rise to domain walls whose lifetime and decay rate will be inversely and directly proportional to the magnitude of the soft breaking mass parameters, respectively. Consequently, increasing the size of the soft-breaking mass parameters will lead to a faster decay of the domain walls and thus to a shorter lifetime for them. Let us note that the vacuum energy density difference produced by the soft-breaking mass term will depend on the soft-breaking parameter. The soft breaking mass term create a pressure difference between the vacua on either side of the domain walls, implying that when the soft breaking parameter is large enough, the pressure difference will cause the domain walls to collapse \cite{Matsuda:1998ms,Dvali:1995cc,Abel:1995wk,Rai:1992xw,Nakayama:2016gxi,Lazarides:2018aev}. Then the bias potential $\Delta V_{\rm bias}$ needs to be high enough in such a way the DW decay before BBN is settled, that is, $\tau_{\rm DW}\approx\sigma_\omega/\Delta V_{\rm bias}\lesssim 0.1$~sec, with $\sigma_\omega\sim v_\sigma^3$ being the DW tension. 
In addition to this, is must be imposed that the DW decay before the DW domination era begins, that is $\tau_{\rm  DW}\lesssim t_{\rm DW}\sim M_P^2/\sigma_\omega$. For  the case where $\Delta V_{\rm bias}=\mu_{\rm bias}^3v_\sigma$, it follows that
\begin{align}
    \frac{\mu_\sigma}{{\rm TeV}}\,\gtrsim\, {\rm max }\left[10^{-5}\,\left(\frac{v_\sigma}{{\rm TeV}}\right)^{2/3},\, 10^{-5}\,\left(\frac{v_\sigma}{{\rm TeV}}\right)^{5/3}\right]. 
\end{align}
It is worth mentioning that among the decay products, gravitational waves are also part, whose signatures may be probed by the current and future observatories~\cite{Hiramatsu:2013qaa,Saikawa:2017hiv}.  
Detailed studies of the domain wall problem, along the lines of \cite{Sassi:2023cqp,Battye:2020sxy,Chen:2020soj}, require a sophisticated analysis of the resulting thermal potential of our model, which will allow to determine the wall energy per unit surface as well as the vacuum energy density difference. These studies are beyond the scope of this work and will be deferred for a future publication.

On the other hand, the muon and electron anomalous magnetic moments receive
contributions due the virtual exchange of heavy neutral (charged) scalars
and charged (neutral) leptons running in the internal lines of the one loop
vertex diagram. Then, the contributions to the muon and electron anomalous
magnetic moments read
\begin{align}
\Delta a_{\mu } &=\dsum\limits_{j=1}^{3}\frac{\func{Re}\left( \beta
_{2}\vartheta _{2}\right) m_{\mu }^{2}}{8\pi ^{2}}\left[ \left(
R_{H}\right) _{1j}\left( R_{H}\right) _{2j}I_{S}^{\left( \mu \right) }\left(
m_{\Psi _{2}},m_{S_{j}}\right) +\left( R_{A}\right) _{1j}\left( R_{A}\right)
_{2j}I_{A}^{\left( \mu \right) }\left( m_{\Psi _{2}},m_{A_{j}}\right) \right]
\notag \\
&+\dsum\limits_{j=1}^{3}\frac{\func{Re}\left( \varkappa _{2}\varrho
_{2}\right) m_{\mu }^{2}}{8\pi ^{2}}\left[ \left( R_{H}\right)
_{1j}\left( R_{H}\right) _{3j}I_{S}^{\left( \mu \right) }\left( m_{\Psi
_{1}},m_{S_{j}}\right) +\left( R_{A}\right) _{1j}\left( R_{A}\right)
_{3j}I_{A}^{\left( \mu \right) }\left( m_{\Psi _{1}},m_{A_{j}}\right) \right]
\notag \\
&+\dsum\limits_{j=1}^{3}\frac{\func{Re}\left( \varkappa _{2}\varrho
_{2}\right) m_{\mu }m_{\Psi _{1}}}{8\pi ^{2}m_{\varphi _{j}^{\pm
}}^{2}}\left( R_{C}\right) _{1j}\left( R_{C}\right) _{3j}J\left( \frac{%
m_{\Psi _{1}}^{2}}{m_{\varphi _{j}^{\pm }}^{2}}\right)   
-\frac{m_{\mu }^{2}}{24\pi ^{2}}\left( \sum_{j=1}^{3}\frac{\left( \left(
R_{C}\right) _{3j}\right) ^{2}\left( \kappa ^{\dagger }\kappa \right) _{22}}{%
m_{\varphi _{j}^{\pm }}^{2}}+4\sum_{k=1}^{3}\frac{\left( G^{\dagger
}G\right) _{22}}{m_{\chi _{k}^{\pm \pm }}^{2}}\right), 
\end{align}%
and
\begin{align}
\Delta a_{e} &=\dsum\limits_{j=1}^{3}\frac{\func{Re}\left( \beta
_{1}\vartheta _{1}\right) m_{e}^{2}}{8\pi ^{2}}\left[ \left(
R_{H}\right) _{1j}\left( R_{H}\right) _{2j}I_{S}^{\left( e\right) }\left(
m_{\Psi _{2}},m_{S_{j}}\right) +\left( R_{A}\right) _{1j}\left( R_{A}\right)
_{2j}I_{A}^{\left( e\right) }\left( m_{\Psi _{2}},m_{A_{j}}\right) \right]  
\notag \\
&+\dsum\limits_{j=1}^{3}\frac{\func{Re}\left( \varkappa _{1}\varrho
_{1}\right) m_{e}^{2}}{8\pi ^{2}}\left[ \left( R_{H}\right)
_{1j}\left( R_{H}\right) _{3j}I_{S}^{\left( e\right) }\left( m_{\Psi
_{1}},m_{S_{j}}\right) +\left( R_{A}\right) _{1j}\left( R_{A}\right)
_{3j}I_{A}^{\left( e\right) }\left( m_{\Psi _{1}},m_{A_{j}}\right) \right]  
\notag \\
&+\dsum\limits_{j=1}^{3}\frac{\func{Re}\left( \varkappa _{1}\varrho
_{1}\right) m_{e}m_{\Psi _{1}}}{8\pi ^{2}m_{\varphi _{j}^{\pm }}^{2}}%
\left( R_{C}\right) _{1j}\left( R_{C}\right) _{3j}J\left( \frac{m_{\Psi
_{1}}^{2}}{m_{\varphi _{j}^{\pm }}^{2}}\right) 
-\frac{m_{e}^{2}}{24\pi ^{2}}\left( \sum_{j=1}^{3}\frac{\left( \left(
R_{C}\right) _{3j}\right) ^{2}\left( \kappa ^{\dagger }\kappa \right) _{11}}{%
m_{\varphi _{j}^{\pm }}^{2}}+4\sum_{k=1}^{3}\frac{\left( G_{k}^{\dagger
}G_{k}\right) _{11}}{m_{\chi _{k}^{\pm \pm }}^{2}}\right), 
\end{align}%
where%
\begin{eqnarray}
&\beta _{1} =\dsum\limits_{i=1}^{3}x_{i}^{\left( l\right) }\left(
V_{lL}^{\dagger }\right) _{1i},\hspace{0.7cm}\vartheta
_{1}=z_{l}\left( V_{lR}\right) _{21}, \hspace{1cm}
\beta _{2} =\dsum\limits_{i=1}^{3}x_{i}^{\left( l\right) }\left(
V_{lL}^{\dagger }\right) _{2i},\hspace{0.7cm}\vartheta
_{2}=z_{l}\left( V_{lR}\right) _{22}, \\
&\varkappa _{1} =\dsum\limits_{i=1}^{3}w_{i}^{\left( l\right) }\left(
V_{lL}^{\dagger }\right) _{1i},\hspace{0.7cm}\varrho
_{1}=r_{l}\left( V_{lR}\right) _{11}, \hspace{1cm}
\varkappa _{2} =\dsum\limits_{i=1}^{3}w_{i}^{\left( l\right) }\left(
V_{lL}^{\dagger }\right) _{2i},\hspace{0.7cm}\varrho
_{2}=r_{l}\left( V_{lR}\right) _{12},
\end{eqnarray}%
being $m_{S_{j}}$, $m_{A_{j}}$, $m_{\varphi _{j}^{\pm }}$ ($j=1,2,3$), $%
m_{\chi _{k}^{\pm \pm }}^{2}$ and $m_{\Psi _{k}}$, ($k=1,2$) the masses of
the CP even neutral $S_{j}$, CP odd neutral $A_{j}$, electrically charged $%
\varphi _{j}^{\pm }$, doubly charged scalars $\chi _{k}^{\pm \pm }$ and
charged exotic vector like fermions $\Psi _{k}$, respectively. Furthermore,
the $I_{S,A}\left( m_{\Psi },m\right)$ and $J\left( r\right)$ loop functions have the
form~\cite{Diaz:2002uk,Jegerlehner:2009ry,Kelso:2014qka,Lindner:2016bgg,Kowalska:2017iqv,Crivellin:2018qmi,Crivellin:2021rbq} 
\begin{equation}
I_{S,A}^{\left( \mu ,e\right) }\left( m_{\Psi },m\right) =\int_{0}^{1}dx%
\frac{x^{2}\left( 1-x\pm \frac{m_{\Psi }}{m_{\mu ,e}}\right) }{m_{\mu
,e}^{2}x^{2}+\left( m_{\Psi }^{2}-m_{\mu ,e}^{2}\right) x+m^{2}\left(
1-x\right) },\hspace{0.7cm}\hspace{0.7cm}J\left( r\right) =\frac{%
-1+r^{2}-2r\ln r}{\left( r-1\right) ^{3}},
\end{equation}%
where $V_{lL}$ and $V_{lR}$ are the rotation matrices that diagonalize $%
\widetilde{M}_{E}$ according to the relation
\begin{equation}
V_{lL}^{\dagger }\widetilde{M}_{E}V_{lR}=\operatorname{diag}\left( m_{e},m_{\mu },m_{\tau
}\right).
\end{equation}%

\begin{figure}[t]
\centering
\includegraphics[width=7cm, height=5cm]{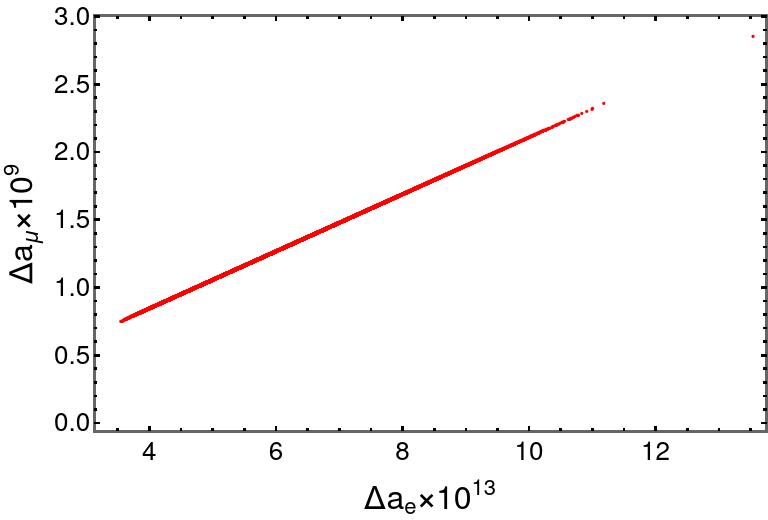}\hspace{1cm}
\includegraphics[width=7cm, height=5cm]{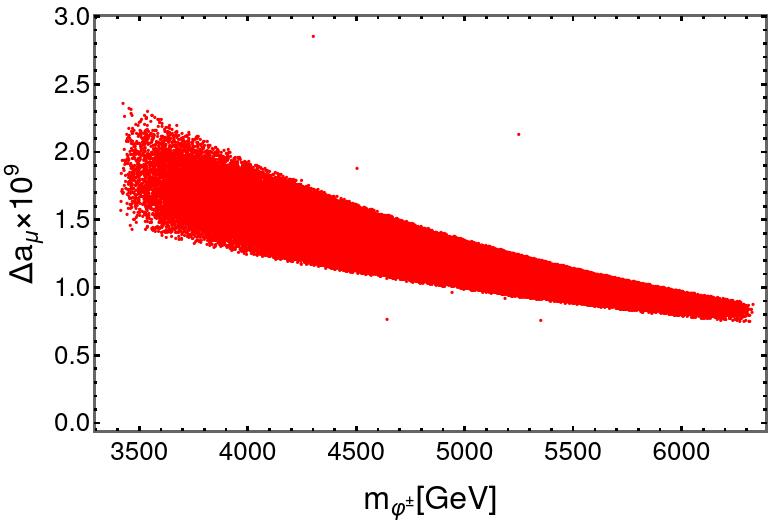}\\
\vspace{0.3cm}
\includegraphics[width=7cm, height=5cm]{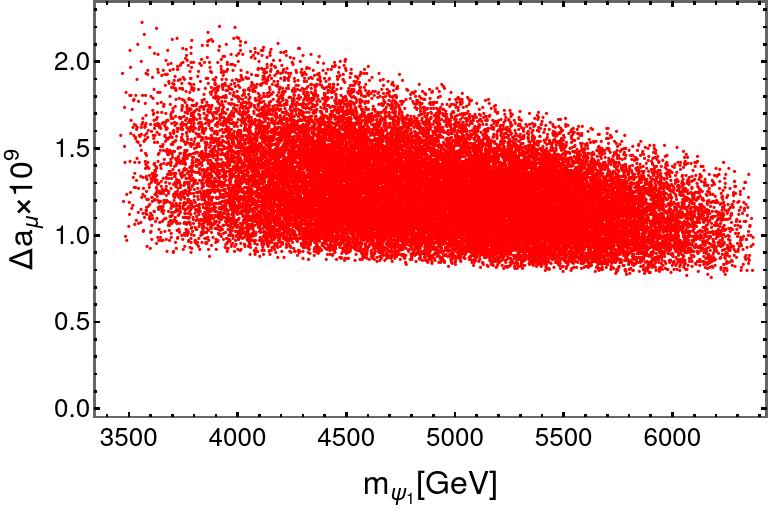}\hspace{1cm}
\includegraphics[width=7cm, height=5cm]{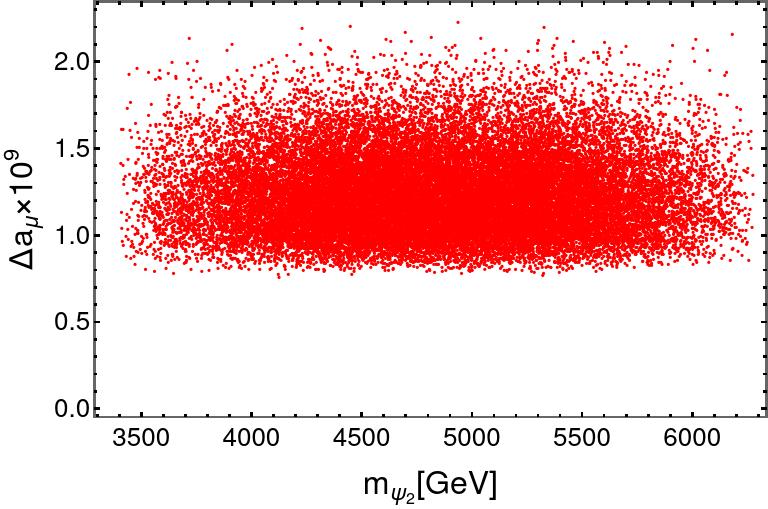}
\caption{Top: correlations between the muon and electron anomalous magnetic
moments (left panel) and of the muon anomalous magnetic moment with the mass $m_{%
\protect\varphi ^{\pm }}$ of the electrically charged scalar $\protect%
\varphi _{1}^{\pm }$ (right panel).
Bottom: correlations of the muon anomalous magnetic moment with the heavy
exotic charged lepton masses $m_{\protect\psi _{1}}$ (left panel) and $m_{\protect\psi %
_{2}}$ (right panel).
}
\label{gminus2}
\end{figure}

\begin{figure}[t]
\centering
\includegraphics[scale=0.3]{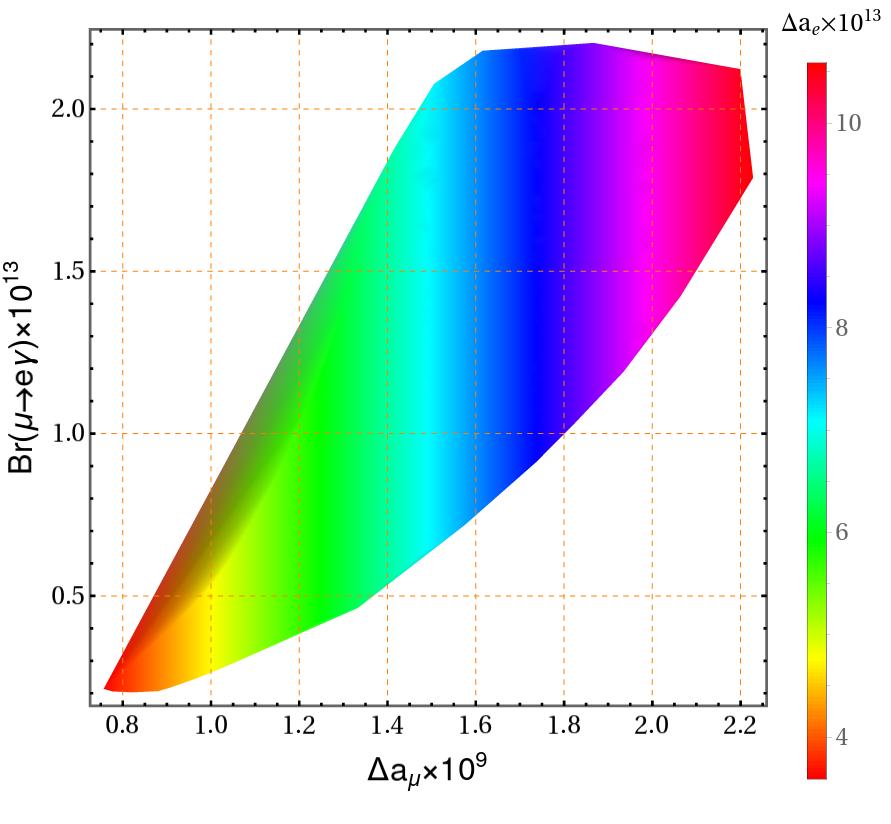}
\caption{Correlation between $Br\left( \protect\mu \rightarrow e\protect%
\gamma \right) $ and the muon anomalous magnetic moment.}
\label{mutoegamma}
\end{figure}

Considering that the muon and electron anomalous magnetic moments are
constrained to be in the ranges \cite{Abi:2021gix,Morel:2020dww}
\begin{eqnarray}
\left( \Delta a_{\mu }\right) _{\exp } &=&\left( 2.51\pm 0.59\right) \times
10^{-9},  \notag \\
(\Delta a_{e})_{\text{exp}} &=&(4.8\pm 3.0)\times 10^{-13}.
\end{eqnarray}%
We plot in Figure \ref{gminus2} (top left panel) the correlation between the electron and
muon anomalous magnetic moments. Furthermore, in this figure are also displayed the correlations of the muon
anomalous magnetic moment with the mass $m_{\varphi ^{\pm }}$ of the
electrically charged scalar $\varphi _{1}^{\pm }$ heavy (top right panel) as well as with the
exotic charged lepton masses $m_{\psi _{1}}$ and $m_{\psi _{2}}$ (bottom panels).
Besides that,
Figure \ref{mutoegamma} displays the correlation between Branching ratio for
the $\mu \rightarrow e\gamma $ decay and the muon anomalous magnetic moment.
To generate these plots, for the sake of simplicity we have considered in
our analysis, the following benchmark scenario: 
\begin{eqnarray}
m_{S_{1}} &=&m_{S},\hspace{0.7cm}m_{S_{2}}=m_{S}+\Delta ,\hspace{0.7cm}%
m_{S_{3}}=m_{S}+2\Delta , \\
m_{A_{1}} &=&m_{A},\hspace{0.7cm}m_{A_{2}}=m_{A}+\Delta ,\hspace{0.7cm}%
m_{A_{3}}=m_{A}+2\Delta , \\
m_{\varphi _{1}^{\pm }} &=&m_{\varphi ^{\pm }},\hspace{0.7cm}m_{\varphi
_{2}^{\pm }}=m_{\varphi ^{\pm }}+\Delta ,\hspace{0.7cm}m_{\varphi _{3}^{\pm
}}=m_{\varphi ^{\pm }}+2\Delta , \\
\Delta &=&100 \;\text{GeV},\hspace{0.7cm}m_{\chi _{k}^{\pm \pm }}=10\;\text{TeV},\hspace{0.7cm}%
k=1,2.
\end{eqnarray}%
and we have varied the masses $m_{S}$, $m_{A}$, $m_{\varphi ^{\pm }}$, $%
m_{\psi _{1}}$ and $m_{\psi _{2}}$ in the ranges 
\begin{equation}
1\;\text{TeV}\leq m_{S},m_{A}\leq 10\;\text{TeV},\hspace{1cm}3\;\text{TeV}\leq
m_{\varphi ^{\pm }}\leq 5\;\text{TeV},\hspace{1cm}1\;\text{TeV}\leq m_{\psi
_{1}},m_{\psi _{2}}\leq 5\;\text{TeV}.
\end{equation}%
As indicated by Figures \ref{gminus2} 
and \ref{mutoegamma}, our model can successfully accommodate the experimental
values of the muon and electron anomalous magnetic moments and is consistent
with the constraints arising from charged lepton flavor violation.
Furthermore, the branching ratio for the $\mu \rightarrow e\gamma $ decay
can reach values of the order of $10^{-13}$, which is within the reach of
future experimental sensitivity, thus making our model testable by the
fourthcoming experiments. In what follows we briefly comment about the implications of the model in the charged lepton flavor violating (CLFV) decays $\mu^- \to e^-e^+e^-$, $\tau^- \to \mu^-\mu^+ \mu^-$, $\tau \to \mu^+ \mu^- e^-$. It is worth mentioming that these CLFV decays take place at tree level thanks to the virtual exchange of the doubly charged scalars of the model. Thus, the experimental upper bounds of these decays~\cite{Bellgardt:1987du} can be used to set constraints on the $\frac{|\gamma_i\gamma^*_j|^2}{m^2_{\chi_k^{\pm\pm}}}$ ($i,j,k=1,2,3$) ratios. For instance, for the $\mu^- \to e^-e^+e^-$ decay, whose branching ratio has the experimental upper bound of $10^{-12}$, one gets the constaint $|\gamma_i\gamma^*_j|<2.3\times 10^{-5}\left(\frac{m_{\chi_j^{\pm\pm}}}{TeV}\right)^2$ \cite{Lindner:2016bgg}, which is successfully fullfilled in our model for appropiate values of the parameters.

Finally, to close this section we provide a brief qualitative discussion about the implications of our model in the electric dipole moment of the neutron. As pointed out in Refs. \cite{Jung:2013hka,Logan:2020mdz}, the electric dipole moment of the neutron in multiHiggs doublet models has several sources: i) tree level CP violating scalar exchange, which give rise to four-fermion operators involving the up- and down-type quarks; ii) the CP-violating three-gluon operator, so called Weinberg operator and the Barr-Zee type two-loop diagrams contributing to the electric dipole moment and chromo-electric dipole
moments of the up- and down-type quarks. It is worth mentioning that the first and third sources of the electric dipole moment of the neutron are suppressed by the small values of the light quark masses \cite{Jung:2013hka,Logan:2020mdz}. Thus, we expect that the leading contribution to the electric dipole moment of the neutron in our extended 2HDM theory will arise from the CP violating two loop level self gluon trilinear interaction involving the virtual exchange of electrically charged scalars as well as top and bottom quarks as in Refs. \cite{Jung:2013hka,Logan:2020mdz}. Therefore, the bound on the electric dipole moment of the neutron $|d_e|\leqslant 1.1\times 10^{-29}$\textit{e cm} \cite{ACME:2018yjb} can be used to set constraints on the ratio between CP violating parameter combinations and squared charged scalar masses, as discussed in detail in Ref. \cite{Jung:2013hka}. Recent comprehensive studies of the implications of multiHiggs doublet models in the electric dipole moment of the neutron are performed in Refs. \cite{Logan:2020mdz,Eeg:2019eei,Eeg:2016fsy}. A detailed numerical analysis of the electric dipole moment of the neutron in this model is beyond the scope of the present work and will be done elsewhere.

\section{Meson mixings}
\label{mesons} 
In this section we discuss the implications of our model in
the Flavour Changing Neutral Current (FCNC) interactions in the down type
quark sector. Given that quark Yukawa interactions include two scalar doublets, there will be tree level flavor changing
neutral currents (FCNC) mediated by neutral scalars and pseudoscalars
exchange that will give rise to $K^{0}-\bar{K}^{0}$, $B_{d}^{0}-\bar{B}%
_{d}^{0}$ and $B_{s}^{0}-\bar{B}_{s}^{0}$ meson oscillations, which can be
described by the following effective Hamiltonians: 
\begin{equation}
\mathcal{H}_{\text{eff}}^{\left( K\right) }\mathcal{=}\sum_{j=1}^{3}\kappa
_{j}^{\left( K\right) }\left( \mu \right) \mathcal{O}_{j}^{\left( K\right)
}\left( \mu \right) ,\hspace{1cm}
\mathcal{H}_{\text{eff}}^{\left( B_{d}\right) }\mathcal{=}\sum_{j=1}^{3}\kappa
_{j}^{\left( B_{d}\right) }\left( \mu \right) \mathcal{O}_{j}^{\left(
B_{d}\right) }\left( \mu \right) ,\hspace{1cm}
\mathcal{H}_{\text{eff}}^{\left( B_{s}\right) }\mathcal{=}\sum_{j=1}^{3}\kappa
_{j}^{\left( B_{s}\right) }\left( \mu \right) \mathcal{O}_{j}^{\left(
B_{s}\right) }\left( \mu \right) ,
\end{equation}
Here $\mathcal{O}_{j}^{\left( K\right) }$, $\mathcal{O}_{j}^{\left(
B_{d}\right) }$ and $\mathcal{O}_{1}^{\left( B_{s}\right) }$ corresponds to
four fermion operators generated after integraring out the scalars and
pseudoscalars that mediate the tree level FCNC interactions producing the $%
K^{0}-\bar{K}^{0}$, $B_{d}^{0}-\bar{B}_{d}^{0}$ and $B_{s}^{0}-\bar{B}%
_{s}^{0}$ meson oscillations. These four fermion operators are given by: 
\begin{eqnarray}
\mathcal{O}_{1}^{\left( K\right) } &=&\left( \overline{s}_{R}d_{L}\right)
\left( \overline{s}_{R}d_{L}\right) ,\hspace{0.7cm}\hspace{0.7cm}\mathcal{O}%
_{2}^{\left( K\right) }=\left( \overline{s}_{L}d_{R}\right) \left( \overline{%
s}_{L}d_{R}\right) ,\hspace{0.7cm}\hspace{0.7cm}\mathcal{O}_{3}^{\left(
K\right) }=\left( \overline{s}_{R}d_{L}\right) \left( \overline{s}%
_{L}d_{R}\right) ,  \label{op3f} \\
\mathcal{O}_{1}^{\left( B_{d}\right) } &=&\left( \overline{d}%
_{R}b_{L}\right) \left( \overline{d}_{R}b_{L}\right) ,\hspace{0.7cm}\hspace{%
0.7cm}\mathcal{O}_{2}^{\left( B_{d}\right) }=\left( \overline{d}%
_{L}b_{R}\right) \left( \overline{d}_{L}b_{R}\right) ,\hspace{0.7cm}\hspace{%
0.7cm}\mathcal{O}_{3}^{\left( B_{d}\right) }=\left( \overline{d}%
_{R}b_{L}\right) \left( \overline{d}_{L}b_{R}\right) , \\
\mathcal{O}_{1}^{\left( B_{s}\right) } &=&\left( \overline{s}%
_{R}b_{L}\right) \left( \overline{s}_{R}b_{L}\right) ,\hspace{0.7cm}\hspace{%
0.7cm}\mathcal{O}_{2}^{\left( B_{s}\right) }=\left( \overline{s}%
_{L}b_{R}\right) \left( \overline{s}_{L}b_{R}\right) ,\hspace{0.7cm}\hspace{%
0.7cm}\mathcal{O}_{3}^{\left( B_{s}\right) }=\left( \overline{s}%
_{R}b_{L}\right) \left( \overline{s}_{L}b_{R}\right) ,
\end{eqnarray}

Besides that $\kappa _{j}^{\left( K\right) }$, $\kappa _{j}^{\left(
B_{d}\right) }$ and $\kappa _{j}^{\left( B_{s}\right) }$ ($j=1,2,3$) are the
corresponding Wilson coefficients which are given by:%
\begin{eqnarray}
\kappa _{1}^{\left( K\right) } &=&\frac{x_{h\overline{s}_{R}d_{L}}^{2}}{%
m_{h}^{2}}+\sum_{n=1}^{3}\left( \frac{x_{S_{n}\overline{s}_{R}d_{L}}^{2}}{%
m_{S_{n}}^{2}}-\frac{x_{A_{n}\overline{s}_{R}d_{L}}^{2}}{m_{A_{n}}^{2}}%
,\right) \\
\kappa _{2}^{\left( K\right) } &=&\frac{x_{h\overline{s}_{L}d_{R}}^{2}}{%
m_{h}^{2}}+\sum_{n=1}^{3}\left( \frac{x_{S_{n}\overline{s}_{L}d_{R}}^{2}}{%
m_{S_{n}}^{2}}-\frac{x_{A_{n}\overline{s}_{L}d_{R}}^{2}}{m_{A_{n}}^{2}}%
\right) ,\hspace{0.7cm}\hspace{0.7cm} \\
\kappa _{3}^{\left( K\right) } &=&\frac{x_{h\overline{s}_{R}d_{L}}x_{h%
\overline{s}_{L}d_{R}}}{m_{h}^{2}}+\sum_{n=1}^{3}\left( \frac{x_{S_{n}%
\overline{s}_{R}d_{L}}x_{S_{n}\overline{s}_{L}d_{R}}}{m_{S_{n}}^{2}}-\frac{%
x_{A_{n}\overline{s}_{R}d_{L}}x_{A_{n}\overline{s}_{L}d_{R}}}{m_{A_{n}}^{2}}%
\right) ,
\end{eqnarray}%
\begin{eqnarray}
\kappa _{1}^{\left( B_{d}\right) } &=&\frac{x_{h\overline{d}_{R}b_{L}}^{2}}{%
m_{h}^{2}}+\sum_{n=1}^{3}\left( \frac{x_{H_{n}\overline{d}_{R}b_{L}}^{2}}{%
m_{S_{n}}^{2}}-\frac{x_{A_{n}\overline{d}_{R}b_{L}}^{2}}{m_{A_{n}}^{2}}%
\right) , \\
\kappa _{2}^{\left( B_{d}\right) } &=&\frac{x_{h\overline{d}_{L}b_{R}}^{2}}{%
m_{h}^{2}}+\sum_{n=1}^{3}\left( \frac{x_{S_{n}\overline{d}_{L}b_{R}}^{2}}{%
m_{S_{n}}^{2}}-\frac{x_{A_{n}\overline{d}_{L}b_{R}}^{2}}{m_{A_{n}}^{2}}%
\right) , \\
\kappa _{3}^{\left( B_{d}\right) } &=&\frac{x_{h\overline{d}_{R}b_{L}}x_{h%
\overline{d}_{L}b_{R}}}{m_{h}^{2}}+\sum_{n=1}^{3}\left( \frac{x_{S_{n}%
\overline{d}_{R}b_{L}}x_{S_{n}\overline{d}_{L}b_{R}}}{m_{S_{n}}^{2}}-\frac{%
x_{A_{n}\overline{d}_{R}b_{L}}x_{A_{n}\overline{d}_{L}b_{R}}}{m_{A_{n}}^{2}}%
\right) ,
\end{eqnarray}%
\begin{eqnarray}
\kappa _{1}^{\left( B_{s}\right) } &=&\frac{x_{h\overline{s}_{R}b_{L}}^{2}}{%
m_{h}^{2}}+\sum_{n=1}^{3}\left( \frac{x_{S_{n}\overline{s}_{R}b_{L}}^{2}}{%
m_{S_{n}}^{2}}-\frac{x_{A_{n}\overline{s}_{R}b_{L}}^{2}}{m_{A_{n}}^{2}}%
\right) , \\
\kappa _{2}^{\left( B_{s}\right) } &=&\frac{x_{h\overline{s}_{L}b_{R}}^{2}}{%
m_{h}^{2}}+\sum_{n=1}^{3}\left( \frac{x_{S_{n}\overline{s}_{L}b_{R}}^{2}}{%
m_{S_{n}}^{2}}-\frac{x_{A_{n}\overline{s}_{L}b_{R}}^{2}}{m_{A_{n}}^{2}}%
\right) , \\
\kappa _{3}^{\left( B_{s}\right) } &=&\frac{x_{h\overline{s}_{R}b_{L}}x_{h%
\overline{s}_{L}b_{R}}}{m_{h}^{2}}+\sum_{n=1}^{3}\left( \frac{x_{S_{n}%
\overline{s}_{R}b_{L}}x_{S_{n}\overline{s}_{L}b_{R}}}{m_{S_{n}}^{2}}-\frac{%
x_{A_{n}\overline{s}_{R}b_{L}}x_{A_{n}\overline{s}_{L}b_{R}}}{m_{A_{n}}^{2}}%
\right) ,
\end{eqnarray}%
The $K-\bar{K}$, $B_{d}^{0}-\bar{B}_{d}^{0}$ and $B_{s}^{0}-\bar{B}_{s}^{0}$
meson mass splittings receive contributions due to Standard Model (SM)
interactions as well as contributions arising from new physics (NP). These meson
mass splittings are given by
\begin{equation}
\Delta m_{K}=\Delta m_{K}^{\left( \text{SM}\right) }+\Delta m_{K}^{\left( \text{NP}\right)
},\hspace{1cm}\Delta m_{B_{d}}=\Delta m_{B_{d}}^{\left( \text{SM}\right) }+\Delta
m_{B_{d}}^{\left( \text{NP}\right) },\hspace{1cm}\Delta m_{B_{s}}=\Delta
m_{B_{s}}^{\left( \text{SM}\right) }+\Delta m_{B_{s}}^{\left( \text{NP}\right) },
\label{Deltam}
\end{equation}%
\begin{figure}
\centering
\includegraphics[width=7cm, height=5cm]{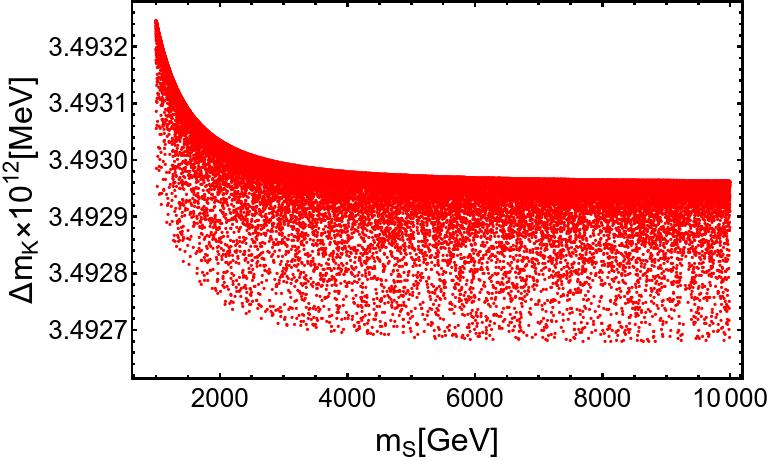}\hspace{0.5cm}
\includegraphics[width=7cm, height=5cm]{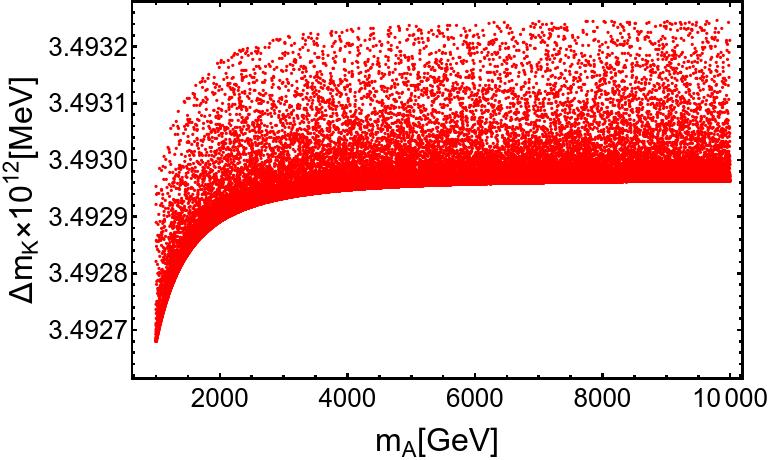}
\caption{Correlation between the $\Delta m_{K}$ mass splitting and the
masses $m_{S}$ ((left panel)) and $m_{A}$ (right panel) of the lightest non SM CP-even and CP-odd
scalars. }
\label{fig:mesonmixing}
\end{figure}
where $\Delta m_{K}^{\left( \text{SM}\right) }$, $\Delta m_{B_{d}}^{\left(
SM\right) }$ and $\Delta m_{B_{s}}^{\left( \text{SM}\right) }$ are the SM
contributions, while $\Delta m_{K}^{\left( \text{NP}\right) }$, $\Delta
m_{B_{d}}^{\left( \text{NP}\right) }$ and $\Delta m_{B_{s}}^{\left( \text{NP}\right) }$
are the contributions arising from tree level flavor changing neutral scalar
interactions. The new physics contributions for the $K-\bar{K}$, $B_{d}^{0}-%
\bar{B}_{d}^{0}$ and $B_{s}^{0}-\bar{B}_{s}^{0}$ meson oscillations obtained
in our model take the form
\begin{equation}
\Delta m_{K}^{\left( \text{NP}\right) }=\frac{8}{3}f_{K}^{2}\eta _{K}B_{K}m_{K}%
\left[ r_{2}^{\left( K\right) }\kappa _{3}^{\left( K\right) }+r_{1}^{\left(
K\right) }\left( \kappa _{1}^{\left( K\right) }+\kappa _{2}^{\left( K\right)
}\right) \right],
\end{equation}%
\begin{equation}
\Delta m_{B_{d}}^{\left( \text{NP}\right) }=\frac{8}{3}f_{B_{d}}^{2}\eta
_{B_{d}}B_{B_{d}}m_{B_{d}}\left[ r_{2}^{\left( B_{d}\right) }\kappa
_{3}^{\left( B_{d}\right) }+r_{1}^{\left( B_{d}\right) }\left( \kappa
_{1}^{\left( B_{d}\right) }+\kappa _{2}^{\left( B_{d}\right) }\right) \right],
\end{equation}%
\begin{equation}
\Delta m_{B_{s}}^{\left( \text{NP}\right) }=\frac{8}{3}f_{B_{s}}^{2}\eta
_{B_{s}}B_{B_{s}}m_{B_{s}}\left[ r_{2}^{\left( B_{s}\right) }\kappa
_{3}^{\left( B_{s}\right) }+r_{1}^{\left( B_{s}\right) }\left( \kappa
_{1}^{\left( B_{s}\right) }+\kappa _{2}^{\left( B_{s}\right) }\right) \right].
\end{equation}%
Using the following numerical values of the meson parameters \cite%
{Dedes:2002er,Aranda:2012bv,Khalil:2013ixa,Queiroz:2016gif,Buras:2016dxz,Ferreira:2017tvy,NguyenTuan:2020xls}%
: 
\begin{eqnarray}
\left( \Delta m_{K}\right) _{\exp } &=&\left( 3.484\pm 0.006\right) \times
10^{-12}\,\text{MeV},\hspace{1.5cm}\left( \Delta m_{K}\right)
_{\text{SM}}=3.483\times 10^{-12}\,\text{MeV}\,,  \notag \\
f_{K} &=&155.7\,\text{MeV}\,,\hspace{1.5cm}B_{K}=0.85,\hspace{1.5cm}\eta
_{K}=0.57\,,  \notag \\
r_{1}^{\left( K\right) } &=&-9.3,\hspace{1.5cm}r_{2}^{\left( K\right) }=30.6,%
\hspace{1.5cm}m_{K}=\left( 497.611\pm 0.013\right) \,\mathrm{{MeV},}
\end{eqnarray}%
\begin{eqnarray}
\left( \Delta m_{B_{d}}\right) _{\exp } &=&\left( 3.334\pm 0.013\right)
\times 10^{-10}\,\text{MeV},\hspace{1.5cm}\left( \Delta m_{B_{d}}\right)
_{\text{SM}}=\left( 3.653\pm 0.037\pm 0.019\right) \times 10^{-10}\,\text{MeV}\,,  \notag
\\
f_{B_{d}} &=&188\,\text{MeV},\hspace{1.5cm}B_{B_{d}}=1.26,\hspace{1.5cm}%
\eta_{B_{d}}=0.55\,,  \notag \\
r_{1}^{\left( B_{d}\right) } &=&-0.52,\hspace{1.5cm}r_{2}^{\left(
B_{d}\right) }=0.88,\hspace{1.5cm}m_{B_{d}}=\left( 5279.65\pm 0.12\right) \,%
\mathrm{{MeV},}
\end{eqnarray}%
\begin{eqnarray}
\left( \Delta m_{B_{s}}\right) _{\exp } &=&\left( 1.1683\pm 0.0013\right)
\times 10^{-8}\,\text{MeV},\hspace{1.5cm}\left( \Delta m_{B_{s}}\right)
_{\text{SM}}=\left( 1.1577\pm 0.022\pm 0.051\right) \times 10^{-8}\,\text{MeV}\,,  \notag
\\
f_{B_{s}} &=&225\,\text{MeV},\hspace{1.5cm}B_{B_{s}}=1.33,\hspace{1.5cm}%
\eta _{B_{s}}=0.55\,,  \notag \\
r_{1}^{\left( B_{s}\right) } &=&-0.52,\hspace{1.5cm}r_{2}^{\left(
B_{s}\right) }=0.88,\hspace{1.5cm}m_{B_{s}}=\left( 5366.9\pm 0.12\right) \,%
\mathrm{{MeV}}.
\end{eqnarray}%
Figure \ref{fig:mesonmixing} displays the correlation between the $\Delta
m_{K}$ mass splitting and the masses $m_{S}$ and $m_{A}$ of the lightest non
SM CP-even and CP-odd scalars. In our numerical analysis, we have considered
the neutral CP even and CP odd scalar masses in the ranges described in the
benchmark scenario chosen in section \ref{gminus2andlfv}. Furthermore, for
the sake of simplicity, we have set the couplings of the flavor changing
neutral Yukawa interactions that produce the $K^{0}-\bar{K}^{0}$
oscillations to be equal to $10^{-6}$. As indicated in Figure \ref%
{fig:mesonmixing}, our model can successfully accommodate the experimental
constraints arising from $K^{0}-\bar{K}^{0}$ meson oscillations. We have
nSIMPs + ZN:umerically checked that the obtained values for the $\Delta m_{B_{d}}$ and $%
\Delta m_{B_{s}}$ mass splittings are also consistent with the experimental
data on meson oscillations for flavor violating Yukawa couplings equal to $%
10^{-4}$ and $5\times 10^{-4}$ for the $B_{d}^{0}-\bar{B}_{d}^{0}$ and $%
B_{s}^{0}-\bar{B}_{s}^{0}$ mixings, respectively.

\section{Conclusions}
\label{conclusions}
We have constructed an extended 2HDM theory with a spontaneously broken $U(1)_{X}$ global symmetry, where the scalar content has been
enlarged by the inclusion of a $SU(2)_{L}$ scalar triplet and
several electrically neutral, charged and doubly charged scalar singlets,
whereas the fermion sector is augmented by adding $SU(2)_{L}$
doublet and singlet charged vector like fermions. The extended particle
content allows the implementation of an extended seesaw mechanism that
generates the first and second generation of the SM charged fermion masses.
In our proposed theory, one $SU(2) $ scalar doublet does acquire a
vacuum expectation value (VEV) at the electroweak symmetry breaking scale
thus generating the top quark mass, whereas the other scalar doublet gets a
VEV of few GeVs thus providing the bottom quark and tau lepton masses. In our
setup, the tiny masses of the light active neutrinos are produced by a two
loop level Zee-Babu mechanism mediated by electrically charged and doubly
charged scalars as well as by SM charged leptons. The model under
consideration is consistent with the current pattern of SM fermion masses
and mixings, with the muon and electron anomalous magnetic moments and
allows to successfully accommodate the constraints arising from charged
lepton flavor violation and meson oscillations. We also have shown that the rate for the charged lepton flavor violating $\mu\to e\gamma$ decay reach values within the reach of the future experimental sensitivity, thus making the model under consideration testable by the forthcoming experiments.

\section*{Acknowledgments}

AECH and IS are supported by ANID-Chile FONDECYT 1210378, 1241855, 1190845, ANID PIA/APOYO
AFB220004, and ANID Programa Milenio code ICN2019$\_$044. The work of DR and
DZ is supported by Sostenibilidad UdeA, UdeA/CODI Grant 2020-33177, and
Minciencias Grants CD 82315 CT ICETEX 2021-1080 and 80740-492-2021. 
O.Z. would like to acknowledge support from the ICTP through the Associates Programme (2023-2028). 
AECH thanks Universidad de Antioquia for hospitality where this work was started.

\bibliographystyle{utphys}
\bibliography{BiblioUS}

\end{document}